\documentclass[10pt,aps,twocolumn,floatfix,showpacs,superscriptaddress]{revtex4-2}
\usepackage{graphicx,amsmath,amsfonts,mathrsfs}
\usepackage{bm}
\usepackage[]{color}

\begin{document}
\title{Exchange interactions and magnetic force theorem}
\author{I.~V.~Solovyev}
\email{SOLOVYEV.Igor@nims.go.jp}
\affiliation{National Institute for Materials Science, MANA,
1-1 Namiki, Tsukuba, Ibaraki 305-0044, Japan}
\affiliation{Department of Theoretical Physics and Applied Mathematics,
Ural Federal University, Mira str. 19, 620002 Ekaterinburg, Russia}
\affiliation{Institute of Metal Physics, S. Kovalevskaya str. 18, 620108 Ekaterinburg, Russia}
\date{\today}

\begin{abstract}
We critically reexamine the problem of interatomic exchange interactions, which describe the total energy change caused by infinitesimal rotations of spins near some equilibrium state in the framework of constrained spin-density functional theory (cSDFT). For the small variations of the spin magnetization, such interactions can be always related to the response function (or transverse spin susceptibility). However, the form of this relation can depend on additional approximations supplementing the practical calculations. Particularly, the commonly used magnetic force theorem prescribes the linear relation between the exchange interactions and the response function, while the exact theory requires this dependence to be inverse, as it can be rigorously derived from cSDFT. We explore the origin and consequences of these differences in the definition for the wide class of materials, including ferromagnetic Ni, antiferromagnetic NiO, half-metallic ferromagnetic CrO$_2$, multiferroic HoMnO$_3$, and layered van der Waals magnets CrCl$_3$ and CrI$_3$. While in most of these cases, the magnetic force theorem produces quite reasonable results and can be rigorously justifies in the long wavelength and strong-coupling limits, the exact formulation appears to be more consistent, especially in dealing with two important issues, which typically arise in the theory of exchange interactions: (i) the treatment of the ligand states, and (ii) the choice of the suitable variable for the description of infinitesimal rotations in the system of spins within cSDFT. Both issues can be efficiently resolved by employing the ideas of adiabatic spin dynamics supplemented with the exact expression for the exchange interactions. Particularly, the ligand states can produce quite sizable contributions to the total energy change. For this case, we propose a simple ``downfolding'' procedure of elimination of the ligand spins from the model by transferring their effect to the interaction parameters between the localized $3d$ spins. Furthermore, the exchange interactions appear to be sensitive to the definition of the variable, which is used in order to describe the rotations of spins in cSDFT: generally, the rotations of spin moments and spin magnetization matrix lead to different results. In this respect, we argue that the rotations of spin moments are more suitable for the description of low-energy excitations, while the rotations of the whole magnetization matrix cause much stronger perturbation in the system of spins.
\end{abstract}
\maketitle

\section{\label{sec:Intro} Introduction}
\par The interatomic exchange interactions is a very useful tool for understanding the properties of magnetic materials on the microscopic level: it is always nice to have a transparent toy model representing a complex magnetic system as a bunch of interacting with each other magnetic centers. Such practice is commonly used in the experiment: for instance, the inelastic neutron scattering data are frequently interpreted in terms of the spin model, which gives us an idea about the main magnetic interactions operating the considered compound. In the theory, the proper spin model can be constructed by eliminating all degrees of freedom except the spin ones, for instance by using perturbation theory~\cite{PWA} or simply mapping the total energy changes obtained for several magnetic configurations onto the spin model, as is frequently done in first-principles electronic structure calculations.

\par Even without spin-orbit coupling, the model can be rather complex and, besides commonly used Heisenberg pair interactions, include other isotropic multispin contributions. There is only a limited number of examples where the simplest Heisenberg form of the model can justified rigorously: (i) The direct exchange interactions, considered by Heisenberg himself~\cite{Heisenberg}; (ii) The strong-coupling limit underlying the superexchange~\cite{PWA} and Ruderman–Kittel–Kasuya–Yosida interactions~\cite{RKKY}; (iii) The effective interactions occurring between infinitesimally rotated spins near some equilibrium state~\cite{LKG1984,LKG1985,LKAG1987}.

\par In the latter case, the Heisenberg form of the model follows from the general property of the 2nd order perturbation theory, which allows us to present energy change caused by local perturbations, occurring at atomic sites, as the sum of pairwise interactions. Furthermore, since without spin-orbit coupling the system is isotropic, these interactions should be described by the scalar products of spins:
\noindent
\begin{equation}
{\cal E} = -\frac{1}{2N} \sum_{ij} J^{ij} \, \boldsymbol{e}_{i} \boldsymbol{e}_{j},
\label{eq:Heis}
\end{equation}
\noindent where $\boldsymbol{e}_{i}$ is the direction of spin at the site $i$ located in the lattice point $\boldsymbol{R}_{i}$, and $N$ is the number of such sites. In the ground state all spins are aligned parallel to $z$. The system of infinitesimally rotated spins can be specified by their transversal components $\delta \boldsymbol{e}_{i}^{\perp} =(\, \cos \boldsymbol{q} \boldsymbol{R}_{i}, \, \sin \boldsymbol{q} \boldsymbol{R}_{i}, \, 0) \, \theta$ for some spin-spiral configuration with the propagation vector $\boldsymbol{q}$. The small polar angle $\theta$ is regarded as a perturbation parameter. Then, if there is only one magnetic site in the unit cell, the energy change caused by interactions of these transversal components of spins is given by $\delta {\cal E} = -\frac{1}{2} J_{\boldsymbol{q}} \theta^{2}$. The generalization to the multi-site case is straightforward and given by:
\noindent
\begin{equation}
\delta {\cal E} = - \frac{1}{2} \sum_{\mu \nu} J_{\boldsymbol{q}}^{\mu \nu} \theta_{\mu} \theta_{\nu},
\label{eq:dexy}
\end{equation}
\noindent with $\mu$ and $\nu$ numbering the atomic sites in the unit cell. This equation contains all necessary information about the interatomic interactions between the spins. The real space parameters $J^{ij}$ can be obtained via the Fourier transform of $J_{\boldsymbol{q}}^{\mu \nu}$. Therefore, the basic idea is to find the corresponding energy change in the electronic structure calculations and map it onto Eq.~(\ref{eq:dexy}). This can be done in the framework of spin-density functional theory (SDFT), which presents a natural way for deriving the parameters of the spin Hamiltonian ``from the first principles''. Eq.~(\ref{eq:dexy}) can be also viewed as the Taylor expansion for the total energy, where each $J_{\boldsymbol{q}}^{\mu \nu}$ is proportional to the 2nd derivative of this energy with respect to $\theta_{\mu}$ and $\theta_{\nu}$, while all 1st derivatives are equal to zero due to the equilibrium condition.

\par The main difficulty on the way of practical realization of this strategy is that it is not always easy to control the rotation of magnetization by the given angles $\theta_{\mu}$ and $\theta_{\nu}$, which should be tuned by applying some external magnetic field. Instead, it is much easier to rotate the exchange-correlation (xc) potential by assuming that within SDFT  it should correspond to rotation of the magnetization by the same angles. This constitutes the basis of the magnetic force theorem (MFT)~\cite{LKG1984,LKG1985,LKAG1987}, which is widely used in practical calculations~\cite{IS2003,Kvashnin,Korotin,Yoon,Nomoto} and was recently extended for treating new exotic magnetic textures~\cite{Grytsiuk}. The great advantage of MFT is that it allows us to replace the total energy change with the change of the single-particle energies~\cite{AFT}. The exchange interactions within MFT are basically given by the transverse susceptibility (or the response function)~\cite{LKG1984,LKG1985,LKAG1987}. Furthermore, starting from this MFT based expression, one can readily reproduce many well-known results for the exchange interactions in the strong-coupling limit~\cite{SClimit}.

\par However, the use of MFT for the exchange interactions is an approximation, which is frequently questioned in the literature~\cite{Stocks,BrunoPRL2003,Antropov}. The exact expression for the exchange interactions is also anticipated and can be related to the \emph{inverse} response function~\cite{BrunoPRL2003,Antropov,Savrasov,Grotheer,KeKatsnelson}. Nevertheless, the issue is still rather controversial as there is no detailed analysis of this problem as well as systematic applications for magnetic materials. The key questions are still: (i) How good is MFT? (ii) Are there any new aspects (besides a quantitative improvement) and/or pitfalls if the exact formalism for the exchange interactions is used instead of MFT?

\par In the present paper, we provide a detailed analysis of this problem starting with the constrained SDFT and focusing on the exact change of the total energy, which corresponds to small rotations of spins near an equilibrium state (Sec.~\ref{sec:Conv}). We will show how the exact expression for the exchange interactions can be derived (Sec.~\ref{sec:exact}) and what are the simplifications underlying the use of MFT (Sec.~\ref{sec:MFT}). Then, we will deal with two important issues, which typically arise in the theory of exchange interactions. The first one is that the real solid consists of several types of states, some of which, like the transition-metal $3d$ states, are primarily responsible for the magnetism, while other ones, like the ligand states, are only magnetised due to the hybridization or weak intraatomic exchange interactions with the $3d$ states and alone would develope no spontaneous magnetization. How is it consistent with the form of the Heisenberg model, which includes only the localized spins? Although the $3d$ states, to certain extent, can be associated with localized spins, the ligand states definitely cannot. To this end, using the adiabaticity concept, we will show how the ligand spins can be naturally eliminated from the model by redefining the magnetic interaction parameters between the $3d$ spins in order to take into account the effect of the ligands (Sec.~\ref{sec:ligand}). Another issue is that the exchange interactions depend on the definition of the object, which is chosen in order to describe the rotations of spins in SDFT (Sec.~\ref{sec:mversusM}). Generally, the rotation of the magnetization matrix does not act the same as the rotation of magnetic moments: these are two different processes, which are characterized by rather different energy scales. The ``right'' choice of such object is still largely phenomenological. Nevertheless, again using the adiabaticity concept, one can argue that the rotation of magnetic moments (instead of the magnetization matrix) should better describe the low-energy excitations in the system of spins. These ideas are illustrated on a number of examples: ferromagnetic (FM) face-centered cubic nickel (fcc Ni, Sec.~\ref{sec:Ni}), antiferromagnetic NiO (Sec.~\ref{sec:NiO}), half-metallic FM CrO$_2$ (Sec.~\ref{sec:CrO2}), multiferroic HoMnO$_3$ (Sec.~\ref{tab:HoMnO3}), and layered van der Waals magnets CrCl$_3$ and CrI$_3$ (Sec.~\ref{sec:CrCl3}). Finally, Sec.~\ref{sec:summary} briefly summarizes results of our work.

\section{\label{sec:Theory} Rotations of magnetization and total energy change}

\subsection{\label{sec:Conv} General conventions and remarks}
\par Our starting point is the constrained SDFT (or its refinements), describing the system of interacting electrons with the energy~\cite{HK,KS}
\noindent
\begin{equation}
{\cal E}[\boldsymbol{m}] = {\cal T}[\boldsymbol{m}] + {\cal E}_{\rm xc}[\boldsymbol{m}] + \frac{1}{2N} \boldsymbol{h}_{\boldsymbol{q}} \cdot \left( \boldsymbol{m} - \boldsymbol{m}_{\boldsymbol{q}} \right),
\label{eq:tenergy}
\end{equation}
\noindent where ${\cal T}$ and ${\cal E}_{\rm xc}$ are, respectively, the kinetic and xc energies (per one unit cell), depending on the spin magnetization $\boldsymbol{m}$, and $\boldsymbol{h}_{\boldsymbol{q}}$ is the constraining field enforcing the given distribution of the spin magnetization $\boldsymbol{m}_{\boldsymbol{q}}$. For the sake of simplicity, we drop here all dependencies on the electron density.

\par The search of the constrained energy in SDFT is reduced to self-consistent solution of one-electron Kohn-Sham (KS) equations with the Hamiltonian $\hat{H}$~\cite{KS}. In order associate the magnetization with the atomic sites, we have to formulate this KS problem on a lattice, by adopting the appropriate representation of localized Wannier orbitals \cite{JPCMreview,WannierRevModPhys} and constructing $\hat{H} = \left[ H_{ij}^{ab}\right]^{\uparrow, \downarrow}$ in the basis of such orbitals, which are denoted as $a$ and $b$ for the atomic sites $i$ and $j$. Furthermore, we assume that the magnetic ground state for $\boldsymbol{h}_{\boldsymbol{q}}=0$ is collinear. Therefore, $\hat{H}$ may depend on the spin indices $\sigma =$ $\uparrow$ or $\downarrow$, but remains diagonal with respect to them. Then, the site-diagonal part of $\hat{H}$ can be presented as $\frac{1}{2} \hat{v}_{i} + \frac{1}{2} \hat{\sigma}^{z} \hat{b}_{i}^{z}$, where $\hat{\boldsymbol{\sigma}} = (\hat{\sigma}^{x}, \hat{\sigma}^{y}, \hat{\sigma}^{z})$ denotes the vector of Pauli matrices, $\hat{v}_{i} = \hat{H}_{ii}^{\uparrow} + \hat{H}_{ii}^{\downarrow}$ represents the scalar potential, and $\hat{b}_{i}^{z} = \hat{H}_{ii}^{\uparrow} - \hat{H}_{ii}^{\downarrow}$ is the xc field, which for an arbitrary direction of the magnetization is given by the vector $\hat{\boldsymbol{b}}_{i} = (\hat{b}_{i}^{x}, \hat{b}_{i}^{y}, \hat{b}_{i}^{z})$.

\par The magnetization at the site $i$ is related to the density matrix
\noindent
\begin{equation}
\hat{n}_{i} = \left(
\begin{array}{cc}
\hat{n}_{i}^{\uparrow   \uparrow} & \hat{n}_{i}^{\uparrow   \downarrow} \\
\hat{n}_{i}^{\downarrow \uparrow} & \hat{n}_{i}^{\downarrow \downarrow}
\end{array}
\right)
\label{eq:dmtrx}
\end{equation}
\noindent as $\hat{\boldsymbol{m}}_{i} = {\rm Tr}_{S} \{ \hat{\boldsymbol{\sigma}} \hat{n}_{i} \}$ (with ${\rm Tr}_{S}$ denoting the trace over the spin indices) and remains a matrix in the subspace spanned by the orbital indices: $\hat{\boldsymbol{m}}_{i} = [\boldsymbol{m}_{i}^{ab}]$. Similar property holds for $\hat{\boldsymbol{b}}_{i}$ and $\hat{\boldsymbol{h}}_{i}$. Hence, the spin moment is given by the trace over the orbital indices: $\boldsymbol{M}_{i} = {\rm Tr}_{L} \left\{ \hat{\boldsymbol{m}}_{i} \right\}$. In SDFT, $\hat{\boldsymbol{b}}_{i}$ is related to $\hat{\boldsymbol{m}}_{i}$ as $\hat{\boldsymbol{b}}_{i} = 2 N \delta {\cal E}_{\rm xc}[\boldsymbol{m}]/\delta \hat{\boldsymbol{m}}_{i}$. If ${\cal E}_{\rm xc}$ is an additive function of $\hat{\boldsymbol{m}}_{i}$ at different sites, $\hat{\boldsymbol{b}}_{i}$ is local and at each site depends only on $\hat{\boldsymbol{m}}_{i}$ at the same site. Then, it is convenient to introduce the vector $\vec{\boldsymbol{m}}^{T} = ( \, \dots \, , \hat{\boldsymbol{m}}_{i}, \, \dots \, )$ composed of $\hat{\boldsymbol{m}}_{i}$ at different sites and similar vectors for the xc and external field: $\vec{\boldsymbol{b}}^{T}$ and $(\vec{\boldsymbol{h}}_{\boldsymbol{q}})^{T}$, respectively.

\par Our goal is to find the energy change caused by infinitesimal rotations of the magnetization near the ground state. Thus, if $\hat{\boldsymbol{m}}_{i} = (0,0,\hat{m}^{z})$ is the translationally invariant ground-state magnetization, the rotated magnetization $\hat{\boldsymbol{m}}_{\boldsymbol{q}i}$ can be written as
\noindent
\begin{equation}
\hat{\boldsymbol{m}}_{\boldsymbol{q}i} = ( \, \theta \cos \boldsymbol{q} \boldsymbol{R}_{i}, \, \theta \sin \boldsymbol{q} \boldsymbol{R}_{i}, \, 1-\frac{\theta^2}{2} ) \, \hat{m}^{z}.
\label{eq:mq}
\end{equation}
\noindent This change of the magnetization is induced by
\noindent
\begin{equation}
\hat{\boldsymbol{h}}_{\boldsymbol{q}i} = (\cos \boldsymbol{q} \boldsymbol{R}_{i}, \sin \boldsymbol{q} \boldsymbol{R}_{i}, 0) \, \hat{h}_{\boldsymbol{q}},
\label{eq:hq}
\end{equation}
\noindent but the angle $\theta$ is additionally affected by the change of the xc field. The corresponding total energy can we written as
\noindent
\begin{widetext}
\begin{equation}
{\cal E}[\vec{\boldsymbol{m}}_{\boldsymbol{q}}] = {\cal E}_{\rm sp} \left( \vec{\boldsymbol{h}}_{\boldsymbol{q}} + \vec{\boldsymbol{b}}_{\boldsymbol{q}} \right) - \frac{1}{2N} \vec{\boldsymbol{m}}_{\boldsymbol{q}} \cdot (\vec{\boldsymbol{h}}_{\boldsymbol{q}} + \vec{\boldsymbol{b}}_{\boldsymbol{q}}) + {\cal E}_{\rm xc}[\vec{\boldsymbol{m}}_{\boldsymbol{q}}],
\label{eq:tenergy2}
\end{equation}
\end{widetext}
\noindent where the first two terms correspond to ${\cal T}$ in Eq.~(\ref{eq:tenergy}): ${\cal E}_{\rm sp}$ is the sum of the occupied KS single-particle energies for the external field $\vec{\boldsymbol{h}}_{\boldsymbol{q}}$ and corresponding to it xc field $\vec{\boldsymbol{b}}_{\boldsymbol{q}}$, while the second term subtracts the interaction of $\vec{\boldsymbol{m}}_{\boldsymbol{q}}$ with these fields. In these notations,  $\vec{\boldsymbol{m}}_{\boldsymbol{q}} \cdot \vec{\boldsymbol{h}}_{\boldsymbol{q}}$ denotes the dot product of two vectors with the summation over two orbital indices as $\sum_{ab} \boldsymbol{m}_{\boldsymbol{q}i}^{ab} \boldsymbol{h}_{\boldsymbol{q}i}^{ba}$ and, if necessary, the atomic indices.

\par Then, ${\cal E}_{\rm xc}[\vec{\boldsymbol{m}}_{\boldsymbol{q}}]$ is invariant with respect to rotations of the spin magnetization, which is a consequence of the gauge invariance in SDFT~\cite{Vignale1987,Vignale1988,PRB1998}. Therefore, ${\cal E}_{\rm xc}[\vec{\boldsymbol{m}}_{\boldsymbol{q}}]$ does not contribute to the total energy change. Similar property holds for $\vec{\boldsymbol{m}}_{\boldsymbol{q}} \cdot \vec{\boldsymbol{b}}_{\boldsymbol{q}}$: due to the gauge invariance, any rotation of the spin magnetization will rotate the xc field by the same amount~\cite{PRB1998}, thus making $\vec{\boldsymbol{m}}_{\boldsymbol{q}} \cdot \vec{\boldsymbol{b}}_{\boldsymbol{q}}$ invariant. This can be clearly seen for the local xc functional of the form
\noindent
\begin{equation}
{\cal E}_{\rm xc}[\vec{\boldsymbol{m}}] = -\frac{1}{4N} \sum_{i} \vec{\boldsymbol{m}}_{i} \cdot \boldsymbol{\mathcal{I}}_{\rm xc} \vec{\boldsymbol{m}}_{i},
\label{eq:exc}
\end{equation}
\noindent where $\boldsymbol{\mathcal{I}}_{\rm xc} = [ \mathcal{I}_{\rm xc}(ab,cd) ]$ is the rank 4 tensor, which can be constructed as discussed in Ref.~\cite{PRB2019}. Then, we have
\noindent
\begin{equation}
\vec{\boldsymbol{b}}_{i} = - \boldsymbol{\mathcal{I}}_{\rm xc} \vec{\boldsymbol{m}}_{i}
\label{eq:bxc}
\end{equation}
\noindent and, therefore, $\vec{\boldsymbol{b}}_{i} \parallel \vec{\boldsymbol{m}}_{i}$. The possibilities other than Eq.~(\ref{eq:exc}) were discussed in Ref.~\cite{PRB1998}. Thus, $\hat{\boldsymbol{m}}_{\boldsymbol{q}i}$ given by Eq.~(\ref{eq:mq}) should correspond to
\noindent
\begin{equation}
\hat{\boldsymbol{b}}_{\boldsymbol{q}i} = ( \, \theta \cos \boldsymbol{q} \boldsymbol{R}_{i}, \, \theta \sin \boldsymbol{q} \boldsymbol{R}_{i}, \, 1-\frac{\theta^2}{2} ) \, \hat{b}^{z},
\label{eq:bq}
\end{equation}
\noindent which consists of transversal, $\delta \vec{\boldsymbol{m}}_{\boldsymbol{q}}^{\perp}$ (i.e., $\in xy$-plane) and longitudinal ($\parallel z$) parts.

\par Then, the change of the single-particle energies is given by~\cite{PRB2014,PRB2019}
\noindent
\begin{widetext}
\begin{equation}
\delta {\cal E}_{\rm sp} = \frac{1}{4N} \left( \vec{\boldsymbol{h}}_{\boldsymbol{q}} + \delta \vec{\boldsymbol{b}}_{\boldsymbol{q}}^{\perp} \right) \cdot \boldsymbol{\mathcal{R}} \left( \vec{\boldsymbol{h}}_{\boldsymbol{q}} + \delta \vec{\boldsymbol{b}}_{\boldsymbol{q}}^{\perp} \right) - \frac{1}{4N} \, \vec{b}^{z} \cdot \vec{m}^{z} \, \theta^2,
\label{eq:spchange1}
\end{equation}
\end{widetext}
\noindent in terms of the rank 4 response tensor $\boldsymbol{\mathcal{R}}$, relating the transversal magnetization, $\delta \vec{\boldsymbol{m}}_{\boldsymbol{q}}^{\perp}$, with the magnetic field: $\boldsymbol{\mathcal{R}} ( \vec{\boldsymbol{h}}_{\boldsymbol{q}} + \delta \vec{\boldsymbol{b}}_{\boldsymbol{q}}^{\perp} ) = \delta \vec{\boldsymbol{m}}_{\boldsymbol{q}}^{\perp}$ (all the details will be given below). The first term in Eq.~(\ref{eq:spchange1}) is nothing but the energy change in the 2nd order of perturbation theory with respect to $\vec{\boldsymbol{h}}_{\boldsymbol{q}} + \delta \vec{\boldsymbol{b}}_{\boldsymbol{q}}^{\perp}$, while the second term appears in the 1st order of perturbation theory with respect to the longitudinal change of the xc field, $-\frac{1}{2} \hat{b}^{z} \theta^2$. Then, using the definition of $\boldsymbol{\mathcal{R}}$ and noting that $\delta \vec{\boldsymbol{b}}_{\boldsymbol{q}}^{\perp} \cdot \delta \vec{\boldsymbol{m}}_{\boldsymbol{q}}^{\perp} = \vec{b}^{z} \cdot \vec{m}^{z} \, \theta^2 $, one can find that
\noindent
\begin{equation}
\delta {\cal E}_{\rm sp} = \frac{1}{4N}  \delta \vec{\boldsymbol{m}}_{\boldsymbol{q}} \cdot  \vec{\boldsymbol{h}}_{\boldsymbol{q}} .
\label{eq:spchange}
\end{equation}
\noindent By combining it with the second term of Eq.~(\ref{eq:tenergy2}) and noting that $\vec{\boldsymbol{m}}_{\boldsymbol{q}}\cdot \vec{\boldsymbol{b}}_{\boldsymbol{q}}$ does not depend on $\theta$, we arrive at simple but exact expression for the total energy change:
\noindent
\begin{equation}
\delta {\cal E} = - \frac{1}{4N} \delta \vec{\boldsymbol{m}}_{\boldsymbol{q}}  \cdot \vec{\boldsymbol{h}}_{\boldsymbol{q}} .
\label{eq:techange}
\end{equation}
\noindent Quite naturally, there would be no energy change without the constraining field.

\subsection{\label{sec:MFT} MFT based expression}
\par Before turning to the exact theory, let us consider the MFT based expression for the exchange interactions. It can be derived from Eq.~(\ref{eq:spchange1}) assuming $\vec{\boldsymbol{h}}_{\boldsymbol{q}} = 0$. In this case, the second term in Eq.~(\ref{eq:tenergy2}) does not contribute to the total energy change, which is formally given only by $\delta {\cal E}_{\rm sp}$. The basic idea here is that $\delta \vec{\boldsymbol{b}}_{\boldsymbol{q}}^{\perp}$ plays the role of constraining field, though it does not guarantee to reproduce the required magnetization change given by Eq.~(\ref{eq:mq}): the input xc field can be indeed taken in the form of Eq.~(\ref{eq:bq}), corresponding to the magnetization (\ref{eq:mq}). However, the new magnetization, obtained from the solution of KS equations with only the xc field (\ref{eq:bq}) will deviated from Eq.~(\ref{eq:mq}) as, without applying the external field, it will tend to relax toward the collinear ground state~\cite{BrunoPRL2003}. This can be paraphrased differently: although for isolated atoms there is one-to-one correspondence between Eqs.~(\ref{eq:mq}) and (\ref{eq:bq}), it is violated in solids because of additional contributions coming from the kinetic energy change, which tend to additionally rotate the magnetization. Thus, the MFT is an approximation. Nevertheless, in many cases it provides quite a reasonable description of the magnetic properties, at least on a semi-quantitative level.

\par Then, it is convenient to make a transformation to the \emph{local coordinate frame}, in which $\delta \hat{\boldsymbol{m}}_{i}^{\perp}$ (and all other vectors) are parallel to $x$: $\delta \hat{\boldsymbol{m}}_{i}^{\perp} = (\delta \hat{m}^{x} , 0, 0)$, and express $\delta \vec{m}^{x}$ via $\delta \vec{b}^{x}$ using the response tensor $\boldsymbol{\mathcal{R}}_{\boldsymbol{q}} \equiv \left[ \mathcal{R}_{ \boldsymbol{q}} (ab , cd) \right]$~\cite{footnote1}:
\noindent
\begin{equation}
\delta \vec{m}^{x} = \boldsymbol{\mathcal{R}}_{\boldsymbol{q}} \left( \vec{h}^{x} + \delta \vec{b}^{x} \right),
\label{eq:mrb}
\end{equation}
\noindent where
\begin{widetext}
\begin{equation}
\mathcal{R}_{ \boldsymbol{q}} (ab , cd) = \frac{1}{2} \sum_{\boldsymbol{k}}^{\rm BZ} \sum_{mn} \frac{f_{m \boldsymbol{k}}^{ \uparrow} - f_{n \boldsymbol{k}+\boldsymbol{q}}^{ \downarrow} }{\varepsilon_{m \boldsymbol{k}}^{ \uparrow} - \varepsilon_{n \boldsymbol{k}+\boldsymbol{q}}^{ \downarrow}} \left\{ (C_{m \boldsymbol{k}}^{a \uparrow})^{*} C_{n \boldsymbol{k}+\boldsymbol{q}}^{b \downarrow} (C_{n \boldsymbol{k}+\boldsymbol{q}}^{c \downarrow})^{*} C_{m \boldsymbol{k}}^{d \uparrow} + {\rm h.c.} \right\},
\label{eq:rabcd}
\end{equation}
\end{widetext}
\noindent in terms of eigenvalues $\varepsilon_{m \boldsymbol{k}}^{ \sigma}$ and eigenvectors $| C_{m \boldsymbol{k}}^{\sigma} \rangle = \left[ \, \dots , \, C_{m \boldsymbol{k}}^{a \sigma}, \, \dots \right]^{T}$ of the KS quasiparticles (in the Bloch representation and expanded in the basis of Wannier orbitals), and the Fermi distribution function $f_{m \boldsymbol{k}}^{ \sigma}$~\cite{PRB2014}. Here, the Hermitian conjugate (h.c.) means the interchange the orbital indices $a \leftrightarrow b$ and $c \leftrightarrow d$ combined with the complex conjugation. The summation over $\boldsymbol{k}$-points runs over the first Brillouin zone (BZ).

\par So far our analysis was limited by one site in the unit cell. The generalization to the multi-site case is straightforward: the elements of the tensor $\boldsymbol{\mathcal{R}}_{\boldsymbol{q}}$ will depend on the indices $\mu$ and $\nu$ of atoms in the unit cell. Furthermore, it should be understood that the orbital indices $a$ and $b$ belong to the site $\mu$, while $c$ and $d$ belong to the site $\nu$.

\par Then, since $\delta \hat{b}_{\mu}^{x} = \theta_{\mu}^{\phantom{z}} \hat{b}_{\mu}^{z}$ and $\vec{h}^{x} = 0$, we will have:
\noindent
\begin{equation}
\delta {\cal E}_{\rm sp} = \frac{1}{4} \sum_{\mu \nu} \left( \vec{b}^{z}_{\mu} \cdot \boldsymbol{\mathcal{R}}_{\boldsymbol{q}}^{\mu \nu} \, \vec{b}^{z}_{\nu} - \vec{b}^{z}_{\mu} \cdot \vec{m}^{z}_{\mu} \, \delta_{\mu \nu}^{\phantom{z}} \right) \theta_{\mu} \theta_{\nu}
\label{eq:mft}
\end{equation}
\noindent and, therefore,
\noindent
\begin{equation}
J_{\boldsymbol{q}}^{\mu \nu} = - \frac{1}{2} \left( \vec{b}^{z}_{\mu} \cdot \boldsymbol{\mathcal{R}}_{\boldsymbol{q}}^{\mu \nu} \, \vec{b}^{z}_{\nu} - \vec{b}^{z}_{\mu} \cdot \vec{m}^{z}_{\mu} \, \delta_{\mu \nu}^{\phantom{z}} \right).
\label{eq:jmft}
\end{equation}
\noindent Taking into account that
\begin{widetext}
\begin{displaymath}
\mathcal{R}_{\boldsymbol{q}}^{\mu \nu} (ab , cd) = -\frac{1}{2 \pi} \sum_{\boldsymbol{k}}^{\rm BZ} {\rm Im} \int_{- \infty}^{\varepsilon_F} d \varepsilon \left\{ G_{\nu \mu}^{da \uparrow}(\varepsilon,\boldsymbol{k}) G_{\mu \nu}^{bc \downarrow}(\varepsilon,\boldsymbol{k}+\boldsymbol{q}) + G_{\nu \mu}^{da \downarrow}(\varepsilon,\boldsymbol{k}+\boldsymbol{q}) G_{\mu \nu}^{bc \uparrow}(\varepsilon,\boldsymbol{k}) \right\},
\end{displaymath}
\end{widetext}
\noindent where $G_{\mu \nu}^{bc \sigma}$ are the matrix elements of the one-electron Green function
\noindent
\begin{displaymath}
\hat{G}^{\sigma}(\varepsilon,\boldsymbol{k}) = \sum_{n} \frac{| C_{n \boldsymbol{k}}^{\sigma} \rangle \langle C_{n \boldsymbol{k}}^{\sigma} | }{\varepsilon - \varepsilon_{n \boldsymbol{k}}^{\sigma} + i \delta},
\end{displaymath}
\noindent it is straightforward to see that Eq.~(\ref{eq:jmft}) is nothing but the MFT based formula for the exchange interactions~\cite{LKG1984,LKG1985,LKAG1987,KL2004}. The second term in the parantheses does not contribute to the real space parameters of interatomic exchange interactions. Nevertheless, it is important in order to fulfil the sum rules. Indeed, using the property
\noindent
\begin{displaymath}
\hat{G}^{\uparrow}(\varepsilon,\boldsymbol{k}) - \hat{G}^{\downarrow}(\varepsilon,\boldsymbol{k}) = \hat{G}^{\uparrow}(\varepsilon,\boldsymbol{k}) \hat{b}^{z} \hat{G}^{\downarrow}(\varepsilon,\boldsymbol{k}),
\end{displaymath}
\noindent which follows from the definition of the Green function, one can find that
\noindent
\begin{displaymath}
\vec{m}^{z}_{\mu} = \sum_{\nu} \boldsymbol{\mathcal{R}}_{0}^{\mu \nu} \, \vec{b}^{z}_{\mu}.
\end{displaymath}
\noindent Then, the second term in Eq.~(\ref{eq:jmft}) can be rearranged as
\noindent
\begin{equation}
\vec{b}^{z}_{\mu} \cdot \vec{m}^{z}_{\mu} = \sum_{\nu} \vec{b}^{z}_{\mu} \cdot \boldsymbol{\mathcal{R}}_{0}^{\mu \nu} \, \vec{b}^{z}_{\nu} = \sum_{\nu} \vec{m}^{z}_{\mu} \cdot \left[ \boldsymbol{\mathcal{R}}_{0}^{-1} \right]^{\mu \nu} \, \vec{m}^{z}_{\nu}.
\label{eq:propertyb}
\end{equation}

\par As was shown in the previous section, the contribution (\ref{eq:mft}) to the total energy change should vanish in the exact formalism due to the cancellation between two contributions in the parentheses. However, in the case of MFT we have $\boldsymbol{\mathcal{R}} \, \delta \vec{\boldsymbol{b}}_{\boldsymbol{q}}^{\perp}  = \delta \vec{\boldsymbol{m}}_{\boldsymbol{q}}'$, which is not the same as the required transversal magnetization $\vec{\boldsymbol{m}}_{\boldsymbol{q}}^{\perp}$. Thus, the cancellation does not occur, but only because of an intrinsic error of MFT for this particular case.

\par The fact that the energy change (\ref{eq:mft}) near the ground state can be fully expressed in terms of the electronic structure of this ground state (thus requiring no additional self-consistency) is regarded as one of the main advantages of MFT~\cite{LKAG1987}. Nevertheless, below we will show that absolutely the same property holds for the exact expression.

\subsection{\label{sec:exact} Exact expression}
\par Now, we turn to the analysis of exact expression (\ref{eq:techange}) for the total energy change. Formally, $\vec{\boldsymbol{h}}_{\boldsymbol{q}}$ serves as an input parameter, while $\delta \vec{\boldsymbol{m}}_{\boldsymbol{q}}^{\perp}$ can be again expressed via $\vec{\boldsymbol{h}}_{\boldsymbol{q}}$ and $\delta \vec{\boldsymbol{b}}_{\boldsymbol{q}}$ using Eq.~(\ref{eq:mrb}) of the linear response theory in the local coordinate frame, which yields
\noindent
\begin{equation}
\delta {\cal E} = -\frac{1}{4} \left( \vec{h}^{x} \cdot \boldsymbol{\mathcal{R}}_{\boldsymbol{q}} \, \vec{h}^{x} + \vec{h}^{x} \cdot \boldsymbol{\mathcal{R}}_{\boldsymbol{q}} \delta \vec{b}^{x} \right),
\label{eq:jexact1}
\end{equation}
\noindent where the dot product implies the summation also over the site indices. However, this expression requires an extra step in order to connect the field $\vec{h}^{x}$ with the angles $\{ \theta_{\mu} \}$. This can be done by using (\ref{eq:mrb}) and applying the self-consistent linear response theory to obtain $\delta \vec{b}^{x}$~\cite{PRB2014,PRB2019,PCCP2019}. Thus, although such procedure can be realized, it is not very practical. It appears to be more convenient to reformulate the problem in a different way, by treating $\delta \vec{\boldsymbol{m}}_{\boldsymbol{q}}^{\perp}$ as the input parameter, and finding corresponding to it $\vec{\boldsymbol{h}}_{\boldsymbol{q}}$ from the linear response theory:
\noindent
\begin{equation}
\vec{h}^{x} = \boldsymbol{\mathcal{R}}_{\boldsymbol{q}}^{-1} \delta \vec{m}^{x} - \delta \vec{b}^{x}.
\label{eq:hrm}
\end{equation}
\noindent which yields:
\noindent
\begin{equation}
J_{\boldsymbol{q}}^{\mu \nu} = \frac{1}{2} \left( \vec{m}^{z}_{\mu} \cdot [ \boldsymbol{\mathcal{R}}_{\boldsymbol{q}}^{-1} ]^{\mu \nu} \, \vec{m}^{z}_{\nu} - \vec{b}^{z}_{\mu} \cdot \vec{m}^{z}_{\mu} \, \delta_{\mu \nu}^{\phantom{z}} \right).
\label{eq:jexact}
\end{equation}
\noindent This expression is an exact analog of Eq.~(\ref{eq:jmft}) and can be formally obtained from it by replacing $\vec{b}^{z}_{\mu} \to \vec{m}^{z}_{\mu}$ and $\boldsymbol{\mathcal{R}}_{\boldsymbol{q}} \to \boldsymbol{\mathcal{R}}_{\boldsymbol{q}}^{-1}$, with the additional change of sign in the whole expression. We would like to emphasize that Eq.~(\ref{eq:jexact}) has the same merits as its MFT-based analog: the exact  interactions are fully determined by the electronic structure and parameters of the ground state. In this sense, the total energy change near the ground state is the property of this ground state, which can be found analytically, without additional self-consistency.

\par Eq.~(\ref{eq:jexact}) can be further rearranged using Eq.~(\ref{eq:bxc}) and expressing $\vec{b}^{z}_{\mu}$ via $\vec{m}^{z}_{\mu}$ as $\vec{b}^{z}_{\mu} = - \boldsymbol{\mathcal{I}}_{\rm xc}^{\mu} \vec{m}^{z}_{\mu}$, which yields
\noindent
\begin{equation}
J_{\boldsymbol{q}}^{\mu \nu} = \frac{1}{2} \left( \vec{m}^{z}_{\mu} \cdot [ \widetilde{\boldsymbol{\mathcal{R}}}_{\boldsymbol{q}}^{-1} ]^{\mu \nu} \, \vec{m}^{z}_{\nu} \right),
\label{eq:jexact2}
\end{equation}
\noindent in terms of the self-consistent response tensor $\widetilde{\boldsymbol{\mathcal{R}}}_{\boldsymbol{q}} = \boldsymbol{\mathcal{R}}_{\boldsymbol{q}} \left[ \bm{\mathsf{1}} + \boldsymbol{\mathcal{I}}_{\rm xc} \boldsymbol{\mathcal{R}}_{\boldsymbol{q}} \right]^{-1}$, satisfying the condition $\delta \vec{m}^{x} = \widetilde{\boldsymbol{\mathcal{R}}}_{\boldsymbol{q}} \, \vec{h}^{x}$.

\par Then, using Eqs.~(\ref{eq:bxc}), (\ref{eq:jmft}), and (\ref{eq:jexact2}), and the approximation $(\vec{b}^{z})^{-1} \otimes (\vec{b}^{z})^{T} \approx \bm{\mathsf{1}}$ for the rank 4 unity tensor $\bm{\mathsf{1}} \equiv [\mathsf{1}(ab,cd)] = [\delta_{ac} \delta_{bd}]$, it is straightforward to obtain the following expression, connecting the exact parameters ($J_{\boldsymbol{q}}$) with the ones based on MFT ($J_{\boldsymbol{q}}^{\tt{MFT}}$):
\noindent
\begin{equation}
J_{\boldsymbol{q}} \approx J_{\boldsymbol{q}}^{\tt{MFT}} \left[ 1 - 2 (\vec{b}^{z})^{-1} \cdot (\vec{m}^{z})^{-1} J_{\boldsymbol{q}}^{\tt{MFT}} \right]^{-1},
\label{eq:RMFT}
\end{equation}
\noindent where we drop for simplicity the atomic indices. In these notations, $(\vec{b}^{z})^{-1}$ is the vector, which for each atomic site is constructed from the elements of the inverse matrix and the dot symbol implies the summation over the orbital indices, as described above. Eq.~(\ref{eq:RMFT}) is nothing but the ``renormalized magnetic force theorem'' proposed by Bruno~\cite{BrunoPRL2003}. Nevertheless, we would like to note here that Bruno considered a spherical case spherical case, where for each atomic site $b^{z}_{ab} = b^{z} \delta_{ab}$ and, therefore, the property $(\vec{b}^{z})^{-1} \otimes (\vec{b}^{z})^{T} = \bm{\mathsf{1}}$ is exact. In a more general case of aspherical $(\vec{b}^{z})$, Eq.~(\ref{eq:RMFT}) is an approximation, while the correct expression is given by Eq.~(\ref{eq:jexact}). Nevertheless, Eq.~(\ref{eq:RMFT}) is very convenient as it shows that $J_{\boldsymbol{q}}$ can be indeed reduced to $J_{\boldsymbol{q}}^{\tt{MFT}}$ at least in two cases: (i) long wavelength limit $\boldsymbol{q} \to 0$ and (ii) strong-coupling limit $\vec{b}^{z} \to \infty$.

\subsection{\label{sec:ligand} Adiabaticity  and elimination of the ligand states}
\par $J_{\boldsymbol{q}}^{\mu \nu}$ contains all the information about the exchange interactions, involving all sites in the unit cell. However, these sites can be of completely different origin. The typical situation is realized in transition-metal (TM) oxides, where the TM $3d$ states are primarily responsible for the magnetism and can be modeled by localized spins. On the other hand, the oxygen sites carry only small magnetic moments, which are induced by the hybridization with the TM $3d$ states. Nonetheless, the magnetic polarization of the oxygen sites plays a very important role by mediating the exchange interactions between the TM sites~\cite{Kanamori_GKA}. Generally, the effect of such polarization is not negligible and should be rigorously taken into account. In the present section we consider how this can be done in the framework of the exact theory of exchange interactions.

\par Let us consider a general situation where all atomic states can be divided in two groups: the magnetic (${\rm T}$) states and the remaining ligand (${\rm L}$) states. Then, the energy change (\ref{eq:dexy}) can be written as:
\noindent
\begin{widetext}
\begin{equation}
\delta {\cal E} = - \frac{1}{2} \left( \, \theta_{\rm T}^{T} J_{\boldsymbol{q}}^{\rm TT} \theta_{\rm T} + \theta_{\rm T}^{T} J_{\boldsymbol{q}}^{\rm TL} \theta_{\rm L} + \theta_{\rm L}^{T} J_{\boldsymbol{q}}^{\rm LT} \theta_{\rm T} + \theta_{\rm L}^{T} J_{\boldsymbol{q}}^{\rm LL} \theta_{\rm L} \right),
\label{eq:eLT}
\end{equation}
\end{widetext}
where each of $J_{\boldsymbol{q}}^{\rm AB}$ is the matrix in the subspace spanned by the indices of T or L states, and $\theta_{\rm A}$ is the column vector with the same indexing (the italic $T$ denotes the matrix transposition, as before).

\par In principle, one can propose several scenarios of how to treat the ${\rm L}$-states. All of them rely on some initial assumptions about the spin dynamics in the system. Namely, the adiabatic spin dynamics implies that all degrees of freedom can be divided into ``slow magnetic'' and ``fast electronic'' ones, so that for each instantaneous configuration of spins, the electronic variables have sufficient time to adjust the magnetic ones and reach the equilibrium~\cite{spindynamics1,spindynamics2}. In this particular case, the key question is what is the nature of the ${\rm L}$-states and whether they should be treated as ``slow'' or ``fast''?~\cite{PRB2019}. Although the question involves many different aspects related to the role of the ${\rm L}$-states and their implications to the magnetic properties of TM compounds, the reasonably good assumption seems to be ``fast''~\cite{PRB2019}, which we will explore below in details.

\par Thus, for each configuration of angles $\theta_{\rm T}$, the angles $\theta_{\rm L}$ can be found from the equilibrium condition: $\frac{\partial} {\partial \theta_{\rm L}^{T}} \delta {\cal E} = 0$, which yields
\noindent
\begin{equation}
\theta_{\rm L} = - \left[ J_{\boldsymbol{q}}^{\rm LL} \right]^{-1} J_{\boldsymbol{q}}^{\rm LT} \theta_{\rm T}.
\label{eq:LfromT}
\end{equation}
\noindent Substituting it into Eq.~(\ref{eq:eLT}) one can eliminate (or downfold) $\theta_{\rm L}$ and obtain the following equation for $\delta {\cal E}$, solely in terms of $\theta_{\rm T}$:
\noindent
\begin{equation}
\delta {\cal E} = - \frac{1}{2} \, \theta_{\rm T}^{T} \, \tilde{J}_{\boldsymbol{q}}^{\rm TT} \, \theta_{\rm T}
\label{eq:eTT}
\end{equation}
\noindent with the downfolded parameters
\noindent
\begin{equation}
\tilde{J}_{\boldsymbol{q}}^{\rm TT} = J_{\boldsymbol{q}}^{\rm TT} -  J_{\boldsymbol{q}}^{\rm TL} \left[ J_{\boldsymbol{q}}^{\rm LL} \right]^{-1} J_{\boldsymbol{q}}^{\rm LT}.
\label{eq:jTT}
\end{equation}

\par This idea of downfolding is quite general and can be applied to any kind of the exchange interactions: exact or approximate ones. However, since it is based on the variational principle and search for the energy minimum for the given configuration of the T-spins, it is more suitable for the exact theory aiming to describe the exact change of the total energy. In this respect, it is important to note that although MFT works reasonably well for the magnetic T-states, the description of the L-states within MFT is more subtle and any attempts to improve MFT (for instance, using Bruno's approach~\cite{BrunoPRL2003}) mainly correct to the behavior of this group of states~\cite{IS2003}. This is also related to the fact that the behavior of the L-states is far from the strong-coupling limit, where MFT is expected to work well. In the view of these arguments, MFT does not seem to be a good starting point for this downfolding procedure and, as we will see below, the exact approach typically produces more consistent results.

\par Finally, we would like to note that a different strategy for the elimination of the L-states has been proposed recently in Ref.~\cite{Logemann}.

\subsection{\label{sec:mversusM} ``Right'' object to rotate: magnetization matrix versus magnetic moments}
\par The next important question is what is the ``right'' perturbation of the spin magnetization at each site of the system, which should be used for the evaluation of the total energy change (\ref{eq:dexy}) and the exchange interaction parameters (\ref{eq:jexact})? One possible answer is $\delta \hat{m}^{x}_{\mu} = \theta_{\mu} \hat{m}^{z}_{\mu}$ (in the local coordinate frame), where each element of the magnetization \emph{matrix} $\hat{m}^{z}_{\mu}$ at the site ${\mu}$ is rotated by the same angle $\theta_{\mu}$. Similar strategy is used in MFT, where the xc field is the \emph{matrix} and each element of this matrix is also rotated by the same angle~\cite{Kvashnin,Korotin}. Nevertheless, such form of the rotation is \emph{our assumption} made about the low-energy excitations in the system of spins, which is materialized in the constraint condition (\ref{eq:tenergy}). Is this choice unique? Are there other perturbations of the spin magnetization matrix, resulting in the same rotations of the spin magnetic moments, but at lower energy cost? In this section, we further explore such possibilities.

\par For the symmetric matrix $\hat{m}^{z}_{\mu}$ one can always choose the diagonal representation $\hat{m}^{z}_{\mu} = {\rm diag} ( \, \dots, \, m^{a,z}_{\mu}, \, \dots )$ with respect to the orbital indices. In principle, each orbital in such representation can be rotated by its own angle $\theta_{\mu}^{a}$, which would yield the transversal magnetization $\hat{m}^{x}_{\mu} = {\rm diag} ( \, \dots, \, \theta_{\mu}^{a} m^{a,z}_{\mu}, \, \dots )$. Nevertheless, these angles are subjected to the additional constraint condition such that the spin moment $M_{\mu}^{x} = {\rm Tr}_{L}\left\{ \hat{m}^{x}_{\mu} \right\}$ should be equal to $\theta_{\mu}^{\phantom{a}} M_{\mu}^{z}$, corresponding to the rotation of $M_{\mu}^{z} = {\rm Tr}_{L}\left\{ \hat{m}^{z}_{\mu} \right\}$ by the angle $\theta_{\mu}^{\phantom{a}}$. Importantly, this condition is softer than rigid rotation of the spin magnetization matrix with the same $\theta_{\mu}^{a} = \theta_{\mu}^{\phantom{a}}$ for all $a$. Therefore, it is reasonable to expect that the energy change will be smaller so as the corresponding exchange coupling parameters. This can be viewed again in the light of the adiabaticity concept, where all degrees of freedom in $\hat{m}^{z}_{\mu}$ are divided in two parts: ``slow'' $M_{\mu}^{z}$ and ``fast'' remaining parameters, which instantaneously follow the rotations of $M_{\mu}^{z}$.

\par The mathematical formulation of the problem should be based on the minimization of the energy change (\ref{eq:techange}) with the additional constraint condition $\sum_{a} \left( \theta^{a}_{\mu} - \theta_{\mu}^{\phantom{a}} \right) m^{a,z}_{\mu} =0$ at each site of the system. Then, this energy change is given by
\noindent
\begin{equation}
\delta {\cal E} = - \frac{1}{4} \sum_{\mu a} \left\{ \theta^{a}_{\mu} m^{a,z}_{\mu} h^{a,x}_{\mu} - \left( \theta^{a}_{\mu} - \theta_{\mu}^{\phantom{a}} \right) m^{a,z}_{\mu} \lambda_{\mu} \right\},
\label{eq:ctheta}
\end{equation}
\noindent where $\lambda_{\mu}$ are the Lagrange multipliers. By minimizing it with respect to $\theta^{a}_{\mu}$ it is straightforward to find that $h^{a,x}_{\mu} = \lambda_{\mu}$. Thus, we arrive at simple but important conclusion: in order to rotate the spin moments at the minimal energy cost, the external field in the subspace of orbital indices at each site should be proportional to the unity matrix, $h^{ab,\, x}_{\mu} = h^{x}_{\mu} \delta_{ab}$.

\par Then, in terms of the linear response theory, we have $\theta_{\mu}^{\phantom{z}} M_{\mu}^{z} = \sum_{\nu} {\rm R}_{\boldsymbol{q}}^{\mu \nu} \left( h^{x}_{\nu} + b^{z}_{\nu} \theta_{\nu}^{\phantom{z}} \right)$, where ${\rm R}_{\boldsymbol{q}}^{\mu \nu} = \sum_{ac} \mathcal{R}_{\boldsymbol{q}}^{\mu \nu} (aa , cc)$ and $b^{z}_{\nu} = \frac{1}{n_{\nu}}{\rm Tr}_{L} \{ \hat{b}^{z}_{\nu} \}$ is the average field at the site $\nu$ (with $n_{\nu}$ being the number of orbitals). The corresponding energy change will be given by Eq.~(\ref{eq:dexy}) with the parameters
\noindent
\begin{equation}
J_{\boldsymbol{q}}^{\mu \nu} = \frac{1}{2} \left( M_{\mu}^{z} \left[{\mathbb R}_{\boldsymbol{q}}^{-1} \right]^{\mu \nu} M_{\nu}^{z} - b^{z}_{\mu} M_{\mu}^{z}\delta_{\mu \nu} \right),
\label{eq:jexactM}
\end{equation}
\noindent where ${\mathbb R}_{\boldsymbol{q}} \equiv [{\rm R}_{\boldsymbol{q}}^{\mu \nu}] $ is the matrix in the subspace of atomic indices. This is an analog of Eq.~(\ref{eq:jexact2}), but reformulated for the rotations of the spin moments instead of the whole magnetization matrix.

\section{\label{sec:appl} Applications}
\par In this section we present results of calculations of the interatomic exchange interactions using the magnetic force theorem, which are denoted as ``$\hat{b}$-based'' (i.e., obtained by rotating the matrix of the xc field), and exact expressions for the energy change corresponding to rotations of the spin magnetization matrix and spin magnetic moments (denoted as ``$\hat{m}$-based'' and ``$M$-based'', respectively). All the calculations were performed using linear muffin-tin orbital (\texttt{LMTO}) method in the atomic spheres approximation~\cite{LMTO1,LMTO2}. Then, for most applications (except fcc Ni) we constructed a minimal model, including only the TM $3d$ and main ligand states. The details will be specified below, separately for each case. The minimal model was constructed in the basis of appropriate Wannier functions by applying the projector operator technique~\cite{WannierRevModPhys,JPCMreview}. We deliberately use the local spin density approximation (LSDA), even despite well known limitations of this approximations for the description of TM oxides and other strongly correlated systems~\cite{AZA}. In this work, we are not aiming at improving LSDA. Nevertheless, we believe that the rigorous analysis of interatomic exchange interactions should shed more light on the problem of what and why should be improved in LSDA. As we will see, in a number of cases the situation can be indeed rather nontrivial. For the practical purposes, we employ the Vosko-Wilk-Nusair LSDA functional~\cite{VWN}.

\subsection{\label{sec:Ni} fcc Ni}
\par The FM fcc Ni is one of the popular testbed systems serving to explore abilities of various theories and models of magnetism~\cite{KL2004}. Therefore, we would also like to start our analysis with the comparison of magnetic interactions in fcc Ni, calculated by employing three different techniques. We use the standard LMTO method in the basis of Ni $3d4sp$ orbitals without the wannierization. Furthermore, the $3d$ states were regarded as ``magnetic states'', while the remaining $4sp$ states were associated with the ``ligand states''. The response tensor was calculated on the mesh of the $90 \times 90 \times 90$ $\boldsymbol{k}$-points and $10 \times 10 \times 10$ $\boldsymbol{q}$-points in the first Brillouin zone.

\par
\begin{figure}[t]
\begin{center}
\includegraphics[width=8.6cm]{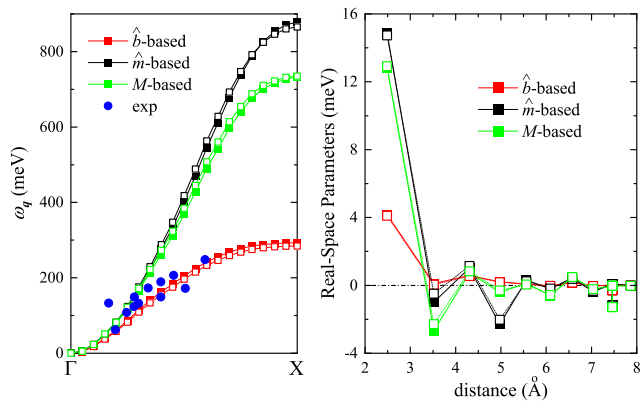}
\end{center}
\caption{Left panel: Spin-wave dispersion for the FM fcc Ni with the parameters obtained in the framework of magnetic force theorem for the infinitesimal rotations of the xc field (denoted as $\hat{b}$-based) and the exact formalism, corresponding to rotations of the whole magnetization matrix ($\hat{m}$-based) and spin magnetic moments ($M$-based). Bare contributions of the Ni $3d$ states are shown by closed symbols, while those taking into account the contributions of the ``ligand'' Ni $4sp$ states are shown by open symbols. The experimental data are from Ref.~\cite{fccNiexp}. Right panel: Distance dependence of interatomic exchange interactions obtained by using the same techniques.}
\label{fig.JNi}
\end{figure}

 The spin-wave dispersion $\omega_{\boldsymbol{q}} = \frac{2}{M} \left( J_{0} - J_{\boldsymbol{q}} \right)$ along the $\Gamma$-${\rm X}$ direction of the Brillouin zone is shown in the left panel of Fig.~\ref{fig.JNi}. The corresponding parameters of exchange interactions in the real space, obtained by the Fourier transform of $J_{\boldsymbol{q}}$, are shown in the right panel. The values of Curie temperature in the mean-field approximation $k_{B} T_{\rm C}^{\rm MF} = \frac{1}{3} \sum_{j} J^{ij}$ (in terms of the real-space parameters $J^{ij}$) and the random phase approximation (RPA)~\cite{tyab},
\begin{equation}
k_{B} T_{\rm C}^{\rm RPA} = \frac{1}{3} \left( \sum_{\boldsymbol{q}} \frac{1}{\left( J_{0} - J_{\boldsymbol{q}} \right)} \right)^{-1},
\label{eq:TRPA}
\end{equation}
\noindent are listed in Table~\ref{tab:TCNi}.
\noindent
\noindent
\begin{table}[b]
\caption{Curie temperature in fcc Ni (in K) as obtained in the mean-field approximation ($T_{\rm C}^{\rm MF}$) and RPA ($T_{\rm C}^{\rm RPA}$) for the parameters derived by rotating (i) the xc field (denoted as $\hat{b}$-based), (ii) the whole magnetization matrix ($\hat{m}$-based), and (iii) spin magnetic moments ($M$-based), where (i) is based on MFT, while (ii) and (iii) are based the exact expression for the total energy change. Bare contributions of the Ni $3d$ states are denoted as $3d$, and the ones including the effect of the ``ligand'' Ni $4sp$ states are denoted as $3d$$+$${\rm L}$. The experimental Curie temperature is about $627$ K~\cite{Skomski}.}
\label{tab:TCNi}
\begin{ruledtabular}
\begin{tabular}{lcccccc}
                        & \multicolumn{2}{c}{$\hat{b}$-based}  & \multicolumn{2}{c}{$\hat{m}$-based}   & \multicolumn{2}{c}{$M$-based} \\
\hline
                        & $3d$     & $3d$$+$${\rm L}$          & $3d$     & $3d$$+$${\rm L}$ & $3d$    & $3d$$+$${\rm L}$  \\
\hline
$T_{\rm C}^{\rm MF}$    & $307$    & $296$                     & $746$    & $760$            & $666$   & $683$            \\
$T_{\rm C}^{\rm RPA}$   & $277$    & $266$                     & $594$    & $611$            & $524$   & $542$
\end{tabular}
\end{ruledtabular}
\end{table}

\par Basically, for fcc Ni we were able to reproduce the main results of Ref.~\cite{KL2004} by Katsnelson and Lichtenstein, which can by summarized as follows: (i) the MFT based exchange parameters better agree with the experimental spin-wave dispersion~\cite{fccNiexp}; (ii) On the other hand, the exact treatment, based on the inverse response function, improves the agreement with the experimental data for $T_{\rm C}$, as was also pointed out by Bruno~\cite{BrunoPRL2003}. Nevertheless, the agreement is merely quantitative as the theoretical values for $T_{\rm C}$ are probably subjected to further corrections including the quantum effects, etc.~\cite{KL2004}. In any case, an interesting point of this analysis is that the simple Heisenberg model with the same parameters fails to describe simultaneously the spin-wave dispersion and $T_{\rm C}$ for fcc Ni, thus confirming results of the previous studies~\cite{spindynamics2}.

\par Regarding the exact theory, in this particular case there is no much difference whether it is formulated in terms of the magnetization matrix ($\hat{m}$-based) or the spin magnetic moments ($M$-based). As expected, the rotations of spin magnetic moments are less energy costly than those of the magnetization matrix. However, in all other respects, these two methods provide quite comparable results for the spin-wave dispersion and the real space parameters of exchange interactions, which substantially exceed the results obtained by rotating the xc field in the framework of MFT. Quite naturally, the magnetism of fcc Ni is almost solely associated with the $3d$ states, while the contributions of the ``ligand'' $4sp$ states are small and do not play a significant role.

\subsection{\label{sec:NiO} Antiferromagnetic NiO}
\par The TM monoxides is another popular class of materials, which is widely used for testing the theories and concepts aiming at the description of strongly correlated systems~\cite{AZA,Oguchi}. A special attention is paid to superexchange interactions responsible for the formation of the type-II antiferromagnetic (AFM) ground state~\cite{Oguchi,PWA,ZaanenSawatzky}. Particularly, the LSDA is known to overestimate these interactions, which is directly related to the underestimation of the energy gap~\cite{AZA,Oguchi}. The main reason is the ``wrong'' averaged interaction parameter $\mathcal{I}_{\rm xc}^{\nu} = \frac{1}{n_{\nu}^{2}} \sum_{ac} \mathcal{I}^{\nu}_{\rm xc}(aa,cc) $, responsible for the splitting between occupied and unoccupied states in LSDA, which should be replaced by much stronger Coulomb repulsion, $U^{\nu}$, enforcing the strong-coupling limit~\cite{AZA}. In this section, we will turn to the analysis of NiO, also within LSDA. Particularly, we will show that in this case MFT substantially overestimates the interatomic exchange interactions and N\'eel temperature, $T_{\rm N}$, in agreement with the previous finding. Nevertheless, the situation is more complex and not only limited to the overestimation of the superexchange interactions. The exact expression, based on the inverse response function,  further deteriorates the agreement with the experimental data.

\par
\begin{figure}[t]
\begin{center}
\includegraphics[width=8.6cm]{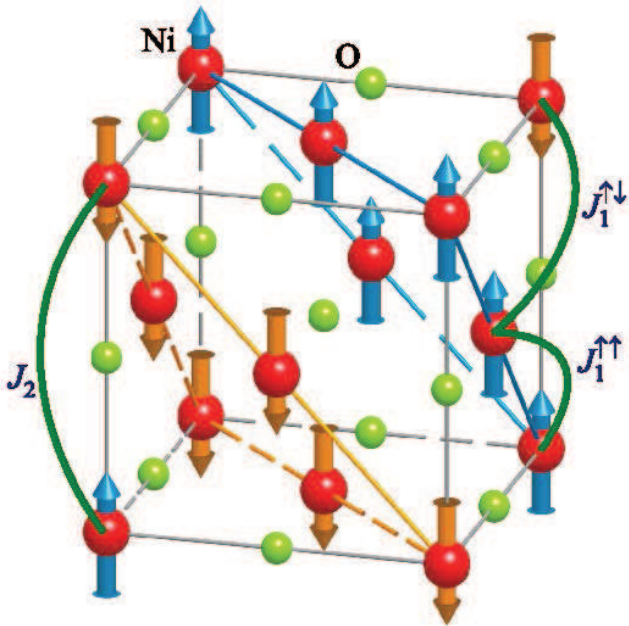}
\end{center}
\caption{Type-II antiferromagnetic phase of NiO with the notation of main exchange interactions.}
\label{fig.NiOstr}
\end{figure}

 We use the minimal model for the electronic structure formulated in the basis of Ni $3d$ and O $2p$ Wannier functions. All calculations are performed for the type-II AFM state in the ideal rock-salt structure (Fig.~\ref{fig.NiOstr}). The response tensor was calculated on the mesh of the $16 \times 16 \times 16$ $\boldsymbol{k}$-points and $10 \times 10 \times 10$ $\boldsymbol{q}$-points in the first Brillouin zone.
\noindent

\par
\begin{figure}[t]
\begin{center}
\includegraphics[width=8.6cm]{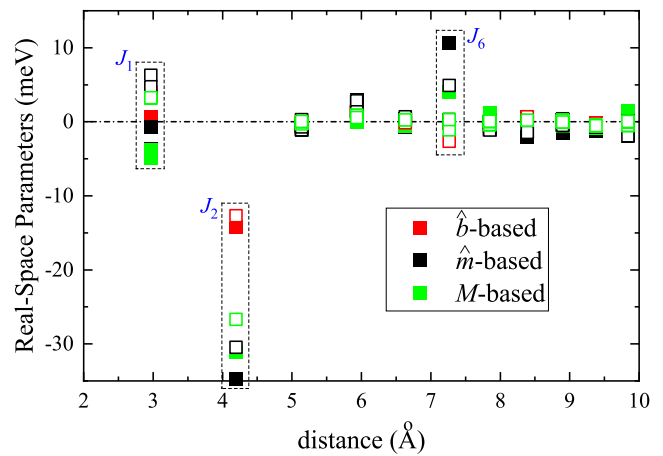}
\end{center}
\caption{Distance dependence of interatomic exchange interactions in NiO as obtained in the framework of MFT for the infinitesimal rotations of the xc field (denoted as $\hat{b}$-based) and the exact formalism, corresponding to rotations of the whole magnetization matrix ($\hat{m}$-based) and only spin magnetic moments ($M$-based). Bare contributions of the Ni $3d$ states are shown by closed symbols. Corrected parameters, taking into account the contributions of the ligand O $2p$ states, are shown by open symbols.}
\label{fig.JNiO}
\end{figure}

 The magnetic properties of NiO are typically considered in terms of the nearest-neighbor (nn) interaction $J_{1}$ and next-nn interaction $J_{2}$, operating via the oxygen sites~\cite{Oguchi,NiOexp} (see Fig.~\ref{fig.NiOstr}). Nevertheless, since LSDA overestimates the itineracy of the system, the exchange interactions in this approximation become long-ranged and not limited by only $J_{1}$ and $J_{2}$. This is clearly seen in Fig.~\ref{fig.JNiO}, illustrating the distance-dependence of exchange interactions: besides $J_{1}$ and $J_{2}$, there is an appreciable interaction $J_{6}$, operating between Ni atoms with opposite directions of spins along the cube diagonal, and other interactions controlling the properties of NiO in LSDA.
\noindent
\noindent However, such long-range behavior is an artifact, resulting from violation of the strong-coupling limit in LSDA. This violation also leads to different values of the parameter $J_{1}$, operating in FM and AFM bonds: $J_{1}^{\uparrow \uparrow}$ and $J_{1}^{\uparrow \downarrow}$, respectively (see Table~\ref{tab:NiO}). The experimental inelastic neutron scattering also indicates at small difference between $J_{1}^{\uparrow \uparrow}$ and $J_{1}^{\uparrow \downarrow}$~\cite{NiOexp}. However, it is much smaller than in LSDA and, more importantly, stems from the small rhombohedral distortion of the rock-salt structure, driven by the exchange striction, while the LSDA parameters correspond to the ideal structure and do not take into account the effect of the distortion.
\noindent
\begin{table}[b]
\caption{Parameters of nearest neighbor and next-nearest neighbor exchange interactions in NiO (in meV) obtained in the framework of magnetic force theorem for the infinitesimal rotations of the xc field (denoted as $\hat{b}$-based) and the exact formalism, corresponding to rotations of the whole magnetization matrix ($\hat{m}$-based) and only spin magnetic moments ($M$-based). Bare contributions of the Ni $3d$ states are denoted as $3d$ and the ones taking into account the effect of the ligand O $2p$ states are denoted as $3d$$+$${\rm L}$. Notations of parameters are explained in Fig.~\ref{fig.NiOstr}. $T_{\rm N}$ is the N\'eel temperature (in K) evaluated within random phase approximation using the complete set of exchange interactions as shown in Fig.~\ref{fig.JNiO}. The experimental parameters are $J_{1}^{\uparrow \uparrow} = 1.39$, $J_{1}^{\uparrow \downarrow} = 1.35$, $J_{2}^{\phantom{\uparrow}} = -19.01$ (all are in meV) and $T_{\rm N} = 523$ K~\cite{NiOexp}.}
\label{tab:NiO}
\begin{ruledtabular}
\begin{tabular}{crrrrrr}
                              & \multicolumn{2}{c}{$\hat{b}$-based}    & \multicolumn{2}{c}{$\hat{m}$-based}     & \multicolumn{2}{c}{$M$-based} \\
\hline
                              & $3d \,$   & $3d$$+$${\rm L}$           & $3d \,$   & $3d$$+$${\rm L}$ & $3d \,$  & $3d$$+$${\rm L}$  \\
\hline
$J_{1}^{\uparrow \uparrow}$   & $0.66$    & $3.53$                     & $-0.65$   & $6.28$           & $-3.78$  & $3.20$            \\
$J_{1}^{\uparrow \downarrow}$ & $0.54$    & $4.08$                     & $-3.62$   & $4.75$           & $-4.85$  & $3.27$            \\
$J_{2}^{\phantom{\uparrow}}$  & $-14.18$  & $-12.70$                   & $-34.76$  & $-30.46$         & $-31.08$ & $-26.69$          \\\hline
$T_{\rm N}$                   & $989$     & $962$                      & $1730$    & $1677$           & $1539$   & $1501$
\end{tabular}
\end{ruledtabular}
\end{table}

\par Now, let us discuss the behavior of $J_{2}^{\phantom{\uparrow}}$ in details. First, we note that the parameter $J_{2}^{\phantom{\uparrow}}$, obtained in the framework of MFT, is even weaker than the experimental one. Certainly, this contradicts to the widespread belief that LSDA should overestimate $|J_{2}^{\phantom{\uparrow}}|$ because it does not include the effects of the on-site Coulomb repulsion, which stands in the denominator of superexchange interactions~\cite{PWA}, and therefore should decrease $|J_{2}^{\phantom{\uparrow}}|$. However, the value $J_{2}^{\phantom{\uparrow}}$ in LSDA is not limited by the superexchange processes and includes other contributions beyond the strong-coupling limit, which can be  ferromagnetic. Thus, $|J_{2}^{\phantom{\uparrow}}|$ in LSDA is not necessarily large. If we took only nn and next-nn interactions from Table~\ref{tab:NiO} and evaluated $T_{\rm N}$ in RPA (also including the quantum factor $1$$+$$1/S$ for $S=1$: all details can be found in Supplemental Materials of Ref.~\cite{PRM2019}), we would get $T_{\rm N} \sim 403$-$465$~K, which is even smaller than the experimental value of $523$~K. Nevertheless, if we take into account \emph{all} interactions, as shown in Fig.~\ref{fig.JNiO}, we obtain instead $T_{\rm N} \sim 962$-$989$~K (see Table~\ref{tab:NiO}), which is larger than the experimental value by almost factor 2. Thus, the problem of LSDA description for NiO is not only (and not necessarily) the overestimation of $|J_{2}^{\phantom{\uparrow}}|$. It is more general: the violation of the strong-coupling limit, which leads to unphysical contributions to $J_{2}^{\phantom{\uparrow}}$ and other (long range) magnetic interactions. Furthermore, such analysis strongly depend on the magnetic state. For instance, rather different picture (with unrealistically large $|J_{2}^{\phantom{\uparrow}}|$) was obtained by Oguchi, Terakura, and Williams~\cite{Oguchi}, who considered the infinitesimal rotations of the xc fields in the paramagnetic state, which is \emph{metallic} within LSDA.

\par Anyway, the exact methods, based on the inverse response function, changes the situation significantly. Particularly, $|J_{2}^{\phantom{\uparrow}}|$ substantially increases. This is reflected in the behavior of $T_{\rm N}$, which also increases and exceeds the experimental value even if one considers only $J_{1}$ and $J_{2}$. The longer-range interactions only aggravate the situation so that the experimental $T_{\rm N}$ becomes overestimated by factor 3. We would like to emphasize that all these changes again manifest the violation of the strong-coupling limit where, according to Eq.~(\ref{eq:RMFT}), the exact parameters are expected to be comparable to the ones in MFT. As expected, the $M$-based scheme produces slightly weaker exchange interactions (and smaller $T_{\rm N}$), but generally the $\hat{m}$ and $M$-based data are comparable. The ligand O $2p$ states systematically strengthen the FM contributions by increasing $J_{1}$ and making somewhat weaker the antiferromagnetic $J_{2}$. Especially, in the exact scheme, the bare interactions $J_{1}^{\uparrow \uparrow}$ and $J_{1}^{\uparrow \downarrow}$ are AFM and only the ligand states make them FM, in agreement with the experiment~\cite{NiOexp}.

\subsection{\label{sec:CrO2} Half-metallic ferromagnetic CrO$_2$}
\par CrO$_2$ provides a rare example of half-metallic ferromagnetism realized in stoichiometric TM oxides. It is widely considered in various applications related the spintronics. Furthermore, it is still regarded as one of the best materials ever invented for magnetic recording~\cite{Skomski}. LSDA is belived to be a reasonably good starting point for the analysis of the magnetic properties of CrO$_2$~\cite{Mazin,HMRevModPhys,CrO2ARPES}.

\par As for NiO, we use the minimal model formulated in the basis of Cr $3d$ and O $2p$ Wannier functions. All calculations are performed for the FM state using the experimental rutile structure (the space group $P4_2/mnm$)~\cite{Porta}. The response tensor was calculated on the mesh of the $20 \times 20 \times 32$ $\boldsymbol{k}$-points and $8 \times 8 \times 12$ $\boldsymbol{q}$-points in the first Brillouin zone.

\par
\begin{figure}[t]
\begin{center}
\includegraphics[width=8.6cm]{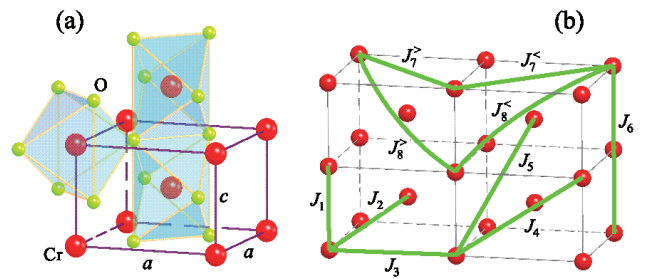}
\end{center}
\caption{(a) Fragment of the crystal structure of CrO$_2$, illustrating the arrangement of the CrO$_6$ octahedra; (b) The lattice of Cr atoms with the notation of the exchange interactions.}
\label{fig.CrO2str}
\end{figure}

 The crystal structure and main magnetic interactions are explained in Fig.~\ref{fig.CrO2str}. The interactions remain sizable up to at least 8th coordination sphere~\cite{PRB2015,JPCM2016}. Moreover, since the rutile structure is nonsymmorphic, there are two types of interactions $J_{7}$ and $J_{8}$, which are denoted by superscripts ``$>$'' and ``$<$''.
\noindent
\noindent The parameters of these interactions, calculated by means of $\hat{b}$-, $\hat{m}$-, and $M$-based techniques, are summarized in Table~\ref{tab:CrO2}, together with the Curie temperature evaluated within RPA.
\noindent
\begin{table}[b]
\caption{Parameters of interatomic exchange interactions in CrO$_2$ (in meV) obtained in the framework of MFT for the infinitesimal rotations of the xc fields (denoted as $\hat{b}$-based) and the exact formalism, corresponding to rotations of the whole magnetization matrix ($\hat{m}$-based) and only spin magnetic moments ($M$-based). Bare contributions of the Cr $3d$ states are denoted as $3d$ and the ones taking into account the effect of the ligand O $2p$ states are denoted as $3d$$+$${\rm L}$. Notations of parameters are explained in Fig.~\ref{fig.CrO2str}. The corresponding Curie temperature ($T_{\rm C}$, in K) is evaluated in RPA.}
\label{tab:CrO2}
\begin{ruledtabular}
\begin{tabular}{crrrrrr}
            & \multicolumn{2}{c}{$\hat{b}$-based}            & \multicolumn{2}{c}{$\hat{m}$-based} & \multicolumn{2}{c}{$M$-based} \\
\hline
            & $3d$$\phantom{-}$ & $3d$$+$${\rm L}$           & $3d$$\phantom{-}$ & $3d$$+$${\rm L}$ & $3d$$\phantom{-}$ & $3d$$+$${\rm L}$  \\
\hline
$J_{1}$     & $30.40$           & $30.66$                    & $45.58$           & $49.07$          & $33.78$           & $37.26$           \\
$J_{2}$     & $20.97$           & $20.96$                    & $26.76$           & $31.24$          & $21.97$           & $24.94$           \\
$J_{3}$     & $2.98$            & $3.05$                     & $2.79$            & $3.97$           & $1.59$            & $2.46$            \\
$J_{4}$     & $1.34$            & $1.36$                     & $0.09$            & $0.02$           & $1.14$            & $1.05$            \\
$J_{5}$     & $-0.82$           & $-0.86$                    & $-1.15$           & $-2.05$          & $-1.04$           & $-1.80$           \\
$J_{6}$     & $-3.58$           & $-3.66$                    & $-3.67$           & $-4.77$          & $-4.29$           & $-5.22$           \\
$J_{7}^{>}$ & $-6.16$           & $-6.21$                    & $-6.82$           & $-9.13$          & $-7.10$           & $-8.48$           \\
$J_{7}^{<}$ & $-1.99$           & $-1.96$                    & $-4.99$           & $-2.89$          & $-4.00$           & $-3.09$           \\
$J_{8}^{>}$ & $-0.55$           & $-0.58$                    & $0.15$            & $-0.73$          & $-0.09$           & $-0.70$           \\
$J_{8}^{<}$ & $-1.39$           & $-1.38$                    & $-3.19$           & $-3.87$          & $-1.59$           & $-1.93$           \\\hline
$T_{\rm C}$ & $820$             & $820$                      & $1016$            & $1215$           & $831$             & $826$
\end{tabular}
\end{ruledtabular}
\end{table}
\noindent All methods predict robust ferromagnetism with $T_{\rm C}$ varying from $820$ to $1215$~K, which substantially exceeds the experimental value of $390$~K~\cite{Skomski}, probably due to neglect of dynamic electron correlations~\cite{PRB2015}.

\par An interesting aspect of CrO$_2$ is the relatively good agreement between results obtained using and the exact method based on the rotation of the spin magnetic moments. As expected, rotations of the magnetization matrix (instead of spin magnetic moments) additionally strengthen the exchange interactions and increase $T_{\rm C}$. However, in this particular case, the effect is not particularly strong. The ligand states do not play a significant role in the MFT based calculations, but become more important in the exact formulism: they increase the FM interactions in the first three coordination spheres. However, this effect is partly compensated by strengthening some AFM interactions in the next coordination spheres, so that $T_{\rm C}$ does not change much.

\subsection{\label{sec:HoMnO3} Multiferroic HoMnO$_3$}
\par In this section we consider capability of different techniques for describing competing exchange interactions, which lead to the breaking of the inversion symmetry in multiferroic manganites with orthorhombic $Pbnm$ structure. We take HoMnO$_3$ as an example. Experimentally, this material displays rather complex magnetic phase diagram. The magnetic transition temperature is about $41$ K. Then, below the so called lock-in transition temperature $T_{\rm L} \approx 29$~K HoMnO$_3$ forms twofold periodic structure with the propagation vector ${\bf k} = (0,\frac{1}{2},0)$, which coincides with the onset of spontaneous ferroelectricity~\cite{ExpStructureHoMnO3,Ishiwata}. The twofold magnetic periodicity is accompanied by the exchange striction and lowering the crystallographic symmetry~\cite{Picozzi,Okuyama}, which we do not consider in the present work. Furthermore, the magnetocrystalline anisotropy can be also important for stabilizing the twofold periodic magnetic texture~\cite{PRB11,PRB12}. Nevertheless, we do not consider these effects either by focusing solely on the behavior of isotropic exchange interactions and the type of the magnetic ground state with the particular direction of ${\bf k}$ along the orthorhombic $\boldsymbol{b}$ axis, while the exchange striction and magnetocrystalline anisotropy are responsible for the particular commensurate value of ${\bf k} = (0,\frac{1}{2},0)$.

\par The details of LMTO calculations can be found in Ref.~\cite{JPSJ}. The calculations have been performed for the (layered) A-type AFM phase using the experimental parameters of the crystal structure reported in Ref.~\cite{ExpStructureHoMnO3}. The minimal model was formulated in the basis of Mn $3d$, O $2p$, and Ho $5d$ Wannier functions. The response tensor was calculated on the mesh of the $14 \times 14 \times 10$ $\boldsymbol{k}$-points and $8 \times 8 \times 6$ $\boldsymbol{q}$-points in the first Brillouin zone.

\par
\begin{figure}[t]
\begin{center}
\includegraphics[width=8.6cm]{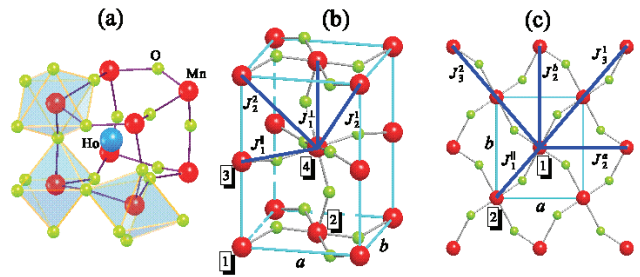}
\end{center}
\caption{(a) Fragment of the crystal structure of HoMnO$_3$, illustrating the arrangement of MnO$_6$ octahedra in the distorted cubic perovskite; (b) the orthorhombic unit cell and (c) the $ab$-plane with the notation of main exchange interactions. Atoms of four Mn sublattices in the primitive cell are denoted by numbers.}
\label{fig.HoMnO3str}
\end{figure}

 Crystal structure of HoMnO$_3$ and main exchange interactions are explained in Fig.~\ref{fig.HoMnO3str}.
\noindent
\noindent Particularly, besides the nn interactions in and between the $\boldsymbol{ab}$ planes ($J_{1}^{\parallel}$ and $J_{1}^{\perp}$, respectively), there are several long-range interactions, which contribute to the properties of HoMnO$_3$ and similar compounds, namely: (i) the next-nn interaction between the planes, $J_{2}^{1}$ and $J_{2}^{2}$; (ii) the 2nd neighbor interactions in the plane, $J_{2}^{a}$ and $J_{2}^{b}$, operating along $\boldsymbol{a}$ and $\boldsymbol{b}$, respectively; and (iii) the 3rd neighbor interactions in the plane, $J_{3}^{1}$ and $J_{3}^{2}$. These interactions obey the symmetry properties of the space group $Pbnm$. For example, around the Mn site 1 in Fig.~\ref{fig.HoMnO3str}b, $J_{2}^{1}$ operates in the bonds $(\frac{a}{2},\pm \frac{b}{2},\frac{c}{2})$ and $(-\frac{a}{2},\pm \frac{b}{2},-\frac{c}{2})$, while $J_{2}^{2}$ operates in the bonds $(-\frac{a}{2},\pm \frac{b}{2},\frac{c}{2})$ and $(\frac{a}{2},\pm \frac{b}{2},-\frac{c}{2})$. The behavior of $J_{2}^{1}$ and $J_{2}^{2}$ around the sites 2, 3, and 4 is obtained by the $180^{\circ}$ rotations of these bonds about $\boldsymbol{a}$, $\boldsymbol{b}$, and $\boldsymbol{c}$ in the combination with the translations by $(\frac{a}{2},\frac{b}{2},0)$, $(0,0,\frac{c}{2})$, and $(\frac{a}{2}, \frac{b}{2}, \frac{c}{2})$, respectively. The same rules can be applied to $J_{3}^{1}$ and $J_{3}^{2}$: around site 1, $J_{3}^{1}$ and $J_{3}^{2}$ operate in the bonds $\pm(a,a,0)$ and $\pm(a,-a,0)$, respectively. The behavior around other sites is obtained by applying above symmetry operations. The interactions $J_{1}^{\perp}$, $J_{2}^{1}$ and $J_{2}^{2}$ are responsible for the AFM coupling between the layers, while the formation of long-periodic magnetic textures in the plane results from the interplay of $J_{1}^{\parallel}$, $J_{2}^{a}$, $J_{2}^{b}$, $J_{3}^{1}$ and $J_{3}^{2}$. The behavior of these interactions is related to the orbital ordering (the preferable population of Mn $3d$ orbitals induced by the cooperative Jahn-Teller distortion)~\cite{PRB12,JPSJ}. The same orbital ordering makes the spin magnetization $\hat{m}$ strongly aspherical.

\par The parameters of exchange interactions are summarized in Table~\ref{tab:HoMnO3}.
\noindent
\begin{table}[t]
\caption{Parameters of interatomic exchange interactions in HoMnO$_3$ (in meV) obtained in the framework of MFT for the infinitesimal rotations of the xc fields (denoted as $\hat{b}$-based) and the exact formalism, corresponding to rotations of the whole magnetization matrix ($\hat{m}$-based) and only spin magnetic moments ($M$-based). Bare contributions of the Mn $3d$ states are denoted as $3d$ and the ones taking into account the effect of the ligand O $2p$ and Ho $5d$ states are denoted as $3d$$+$${\rm L}$. Notations of exchange interactions are explained in Fig.~\ref{fig.HoMnO3str}. $T_{\bf k}$ is the magnetic transition temperature (in K) evaluated in RPA. ${\bf k}$ denotes the magnetic propagation vector.}
\label{tab:HoMnO3}
\begin{ruledtabular}
\begin{tabular}{cccrrcc}
                    & \multicolumn{2}{c}{$\hat{b}$-based}            & \multicolumn{2}{c}{$\hat{m}$-based} & \multicolumn{2}{c}{$M$-based} \\
\hline
                    & $~3d$             & $~3d$$+$${\rm L}$          & $3d$$\phantom{-}$ & $3d$$+$${\rm L}$ & $~3d$             & $~3d$$+$${\rm L}$ \\
\hline
$J_{1}^{\parallel}$ & $\phantom{-}2.78$ & $\phantom{-}0.08$          & $-0.89$           & $27.93$          & $-5.47$           & $\phantom{-}7.12$ \\
$J_{1}^{\perp}$     & $-0.44$           & $-0.15$                    & $-18.26$          & $-15.51$         & $-6.38$           & $-6.27$           \\
$J_{2}^{1}$         & $-0.92$           & $-0.69$                    & $-6.94$           & $-1.90$          & $-4.54$           & $-1.17$           \\
$J_{2}^{2}$         & $-0.88$           & $-0.69$                    & $-7.15$           & $-2.33$          & $-4.07$           & $-1.27$           \\
$J_{2}^{a}$         & $\phantom{-}2.63$ & $\phantom{-}1.74$          & $-32.85$          & $-7.48$          & $-3.92$           & $\phantom{-}1.00$ \\
$J_{2}^{b}$         & $-1.50$           & $ -1.18$                   & $-14.50$          & $-4.62$          & $-5.55$           & $-1.01$           \\
$J_{3}^{1}$         & $\phantom{-}1.38$ & $\phantom{-}1.32$          & $-6.78$           & $-13.93$         & $-9.27$           & $-6.35$           \\
$J_{3}^{2}$         & $\phantom{-}3.38$ & $\phantom{-}2.22$          & $-18.93$          & $-2.83$          & $-0.87$           & $-0.17$           \\\hline
$T_{\bf k}$         & $~119$            & $~54$                      & $381~$            & $235~$           & $~110$            & $~82$            \\
type                & $\, \, {\bf k}$$=$$0$   & $\, \, {\bf k}$$=$$0$            & ${\bf k}$$\perp$$\boldsymbol{b} \,$        & ${\bf k}$$ \perp$$\boldsymbol{b} \,$       & $\, \,{\bf k}$$\parallel$$\boldsymbol{b}$       & $\, \, {\bf k}$$\parallel$$\boldsymbol{b}$
\end{tabular}
\end{ruledtabular}
\end{table}
\noindent All techniques correctly reproduce the AFM coupling between the planes. Nevertheless, there is a substantial difference in the behavior of magnetic interactions within the plane. Particularly, the magnetic force theorem predicts $J_{2}^{a}$ and $J_{2}^{b}$ to be FM and AFM, respectively, which is consistent with the twofold periodicity along $\boldsymbol{b}$. Nevertheless, the interactions $J_{3}^{1}$ and $J_{3}^{2}$ are FM and stronger than $J_{2}^{b}$. Therefore, the symmetry breaking does not occur and the system remains in the A-type AFM state. Note that the long-range interactions in LSDA are expected to be strongly oscillating~\cite{Heine} and can easily change sign depending on the method used for their calculations.

\par In the exact $\hat{m}$-based method, all interactions $J_{2}$ and $J_{3}$ are AFM. However, $J_{2}^{b}$ appears to be weaker than $J_{2}^{a}$. Then, the magnetic symmetry breaking does occur, but the propagation vector ${\bf k}$ is perpendicular to $\boldsymbol{b}$. Moreover, like in other applications of the $\hat{m}$-based technique, the exchange interactions and the magnetic transition temperature ($T_{\bf k}$) are strongly overestimated. Apparently, such discrepancy is related to the strong asphericity of $\hat{m}$, and the rotations of $\hat{m}$, which preserve this asphericity, do not describe properly (neither quantitatively nor even qualitatively) the energy change associated with the small rotations of spins in HoMnO$_3$.

\par It appears that the only technique, which correctly reproduces the type of the magnetic ground state and the direction of ${\bf k}$ in HoMnO$_3$, is $M$-based (i.e., rotating the spin magnetic moments instead of the whole magnetization matrix). In this case, $J_{2}^{b}$ is stronger than $J_{2}^{a}$ and all $J_{3}$ are AFM, yielding the incommensurate magnetic ground state with ${\bf k} \parallel \boldsymbol{b}$. The ligand states mainly affect the quantitative estimates, while the main tendencies are reproduced by bare exchange interactions between the Mn $3d$ states. For instance, ${\bf k}$ changes from $(0,0.46,0)$ in the bare case till $(0,0.30,0)$ when the ligand states are taken into account. Moreover, the ligand states somewhat decrease $T_{\bf k}$ (see Table~\ref{tab:HoMnO3}). The magnetic transition temperature is overestimated by factor 2~\cite{ExpStructureHoMnO3,Ishiwata}: partly because of the limitations of LSDA, partly because of the oversimplification of the problem and neglect of other important ingredients, which lead to the realization of experimental incommensurate sinusoidal spin structure just below the transition temperature.

\subsection{\label{sec:CrCl3} Layered Chromium Trihalides}
\par Chromium trihalides, Cr$X_3$ ($X$$=$ Cr or I), have attracted a considerable attention as candidates in the search for magnetic two-dimensional materials, which could be important for developing ultracompact spintronic devices~\cite{CrI3_Nature}. Indeed, these materials form a layered van der Waals structure and, therefore, can be rather easily prepared in the two-dimensional form.

\par The details of LMTO calculations can be found in Ref.~\cite{PRB2019}. These calculations have been performed for the FM state using experimental parameters of the $R \overline{3}$ structure reported in Refs.~\cite{CrCl3str,CrI3str}. The minimal model was formulated in the basis of Cr $3d$ and Cr $3p$ (I $5p$) Wannier functions. The response tensor was calculated on the mesh of the $10 \times 10 \times 10$ $\boldsymbol{k}$-points and the same mesh of $\boldsymbol{q}$-points in the first Brillouin zone.

\par
\begin{figure}[b]
\begin{center}
\includegraphics[width=8.6cm]{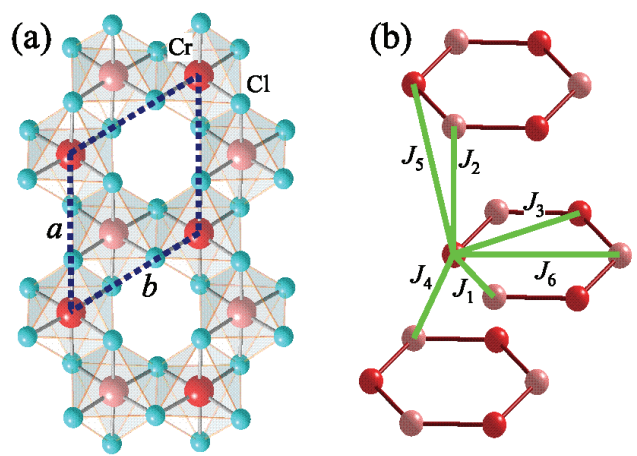}
\end{center}
\caption{(a) Top view on the CrCl$_3$ layer. The unit cell is denoted by the broken line. (b) Stacking of adjacent layers with the notation of main exchange interactions. Two Cr sites in the unit cell are denoted by different colors.}
\label{fig.CrCl3str}
\end{figure}

\begin{figure}[t]
\begin{center}
\includegraphics[width=8.6cm]{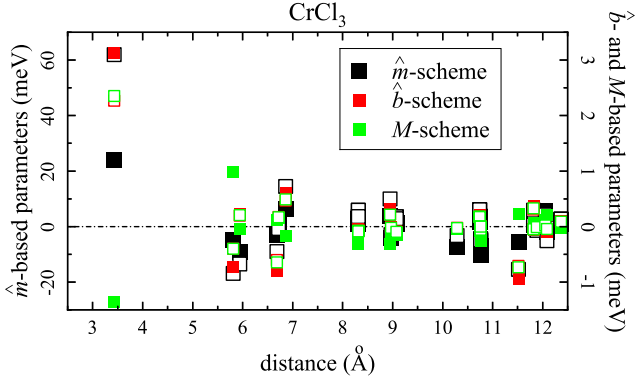}
\end{center}
\caption{Distance dependence of interatomic exchange interactions in CrCl$_3$ as obtained in the framework of magnetic force theorem for the rotations of the xc field (denoted as $\hat{b}$-based) and using the exact expression for the total energy change corresponding to rotations of the whole magnetization matrix ($\hat{m}$-based) and spin magnetic moments ($M$-based). Bare contributions of the Cr $3d$ states are shown by closed symbols. Corrected parameters, which include the contributions of the ligand Cl $3p$ states, are shown by open symbols. Note different $y$-axis scale used for the $\hat{m}$-based parameters (left) and $\hat{b}$- and $M$-based parameters (right).}
\label{fig.JCrCl3}
\end{figure}

 Crystal structure of CrCl$_3$ and main exchange interactions are explained in Fig.~\ref{fig.CrCl3str} and the distance-dependence of these interactions is shown in Fig.~\ref{fig.JCrCl3}.
\noindent
\noindent We note sizable interactions spreading at least up to 6th nearest neighbors at the distance $\sim 6.9$~\AA and beyond: for instance, there is a strong interaction between Cr sites separated by the hexagonal translation $(0,0,c)$ at the distance $\sim 11.5$~\AA, etc.
\noindent

\par The exchange parameters, evaluated using three different techniques, are summarized in Tables~\ref{tab:CrCl} and \ref{tab:CrI}, for CrCl$_3$ and CrI$_3$, respectively.
\noindent
\begin{table}[b]
\caption{Parameters of interatomic exchange interactions in CrCl$_3$ (in meV) obtained by rotating the xc field in the framework of MFT (denoted as $\hat{b}$-based) and using on the exact energy change, corresponding to rotations of the whole magnetization matrix ($\hat{m}$-based) and spin magnetic moments ($M$-based). Bare contributions of the Cr $3d$ states are denoted as $3d$ and the ones taking into account the effects of the ligand Cl $3p$ states are denoted as $3d$$+$${\rm L}$. Notations of exchange parameters are explained in Fig.~\ref{fig.CrCl3str}. $T_{\rm X}$ is corresponding magnetic transition temperature (in K) evaluated in RPA. The type of the magnetic ground state (${\rm X}$) is discussed in the text.}
\label{tab:CrCl}
\begin{ruledtabular}
\begin{tabular}{crrrrrr}
                              & \multicolumn{2}{c}{$\hat{b}$-based}    & \multicolumn{2}{c}{$\hat{m}$-based}     & \multicolumn{2}{c}{$M$-based} \\
\hline
            & $3d \,$ & $3d$$+$${\rm L}$ & $3d \,$  & $3d$$+$${\rm L}$ & $3d \,$  & $3d$$+$${\rm L}$  \\
\hline
$J_{1}$     & $3.12$  & $2.27$           & $23.98$  & $61.93$          & $-1.36$  & $2.35$            \\
$J_{2}$     & $-0.72$ & $-0.40$          & $-4.93$  & $-16.90$         & $0.98$   & $-0.40$           \\
$J_{3}$     & $0.23$  & $0.20$           & $-9.11$  & $-13.53$         & $-0.05$  & $0.21$            \\
$J_{4}$     & $-0.80$ & $-0.60$          & $-3.11$  & $-9.03$          & $0.13$   & $-0.64$           \\
$J_{5}$     & $0.15$  & $0.17$           & $0.60$   & $-0.67$          & $0.16$   & $0.17$            \\
$J_{6}$     & $0.61$  & $0.47$           & $6.31$   & $14.48$          & $-0.17$  & $0.49$            \\\hline
$T_{\rm X}$ & $55 \,$ & $47 \,$          & $699 \,$ & $486 \,$         & $22 \,$  & $50 \,$
\end{tabular}
\end{ruledtabular}
\end{table}
\noindent
\begin{table}[t]
\caption{The same as Table~\ref{tab:CrCl} but for CrI$_3$.}
\label{tab:CrI}
\begin{ruledtabular}
\begin{tabular}{crrrrrr}
                              & \multicolumn{2}{c}{$\hat{b}$-based}    & \multicolumn{2}{c}{$\hat{m}$-based}     & \multicolumn{2}{c}{$M$-based} \\
\hline
            & $3d \,$ & $3d$$+$${\rm L}$ & $3d \,$  & $3d$$+$${\rm L}$ & $3d \,$  & $3d$$+$${\rm L}$  \\
\hline
$J_{1}$     & $1.97$  & $1.00$           & $11.65$  & $51.82$          & $-9.32$  & $1.38$            \\
$J_{2}$     & $0.07$  & $0.40$           & $-1.74$  & $-12.22$         & $2.86$   & $0.24$            \\
$J_{3}$     & $0.80$  & $0.81$           & $-1.74$  & $-8.46$          & $-0.60$  & $0.77$            \\
$J_{4}$     & $-0.17$ & $0.06$           & $-1.32$  & $-7.68$          & $1.80$   & $-0.05$           \\
$J_{5}$     & $0.60$  & $0.62$           & $1.24$   & $-0.93$          & $-0.05$  & $0.59$            \\
$J_{6}$     & $0.43$  & $0.27$           & $3.07$   & $11.63$          & $-2.03$  & $0.30$            \\\hline
$T_{\rm X}$ & $91 \,$ & $67 \,$          & $652 \,$ & $499 \,$         & $132 \,$ & $67 \,$
\end{tabular}
\end{ruledtabular}
\end{table}
\noindent An interesting aspect of these materials is that the obtained exchange interactions strongly depend on the method of their calculation. From this point of view, these systems are particularly interesting for the purposes of our work. Let us consider first the nn interaction in the honeycomb plane, $J_{1}$, which for the nearly $90^\circ$ exchange path Cr-$X$-Cr is expected to be ferromagnetic according to the Goodenough-Kanamori-Anderson rules ~\cite{Kanamori_GKA}. However, this ferromagnetism arises mainly from to the intraatomic (Hund's rule) exchange interaction at the ligand sites so that the result strongly depends on whether and how these interactions are included to the particular scheme for the evaluation of interatomic exchange integrals~\cite{PRB2019}. Indeed, MFT predicts bare $J_{1}$ to be weakly ferromagnetic, in both CrCl$_3$ and CrI$_3$. Rather counterintuitively, the ligand states decrease $J_{1}$. On the contrary, the $\hat{m}$-scheme yields the robust FM coupling $J_{1}$, which is strongly enhanced by the ligand states. Nevertheless, it does not mean that the ground state is ferromagnetic and in a moment we will see that (also strong) longer-range AFM interactions in the $\hat{m}$-case lead to a non-collinear magnetic alignment. Then, in the $M$-scheme, the bare exchange integral $J_{1}$ is antiferromagnetic. This coupling is relatively weak in CrCl$_3$, but becomes strong in CrI$_3$. However, the ligand states change the situation dramatically and restores the ferromagnetism, as it should be. Thus, in the $M$-scheme, the FM character of the coupling $J_{1}$ is entirely related to the ligand states. One may also ask why the bare $J_{1}$ is so different in the $\hat{b}$-, $\hat{m}$-, and $M$-based methods? In fact, the bare $J_{1}$ includes several different contributions. For instance, considering only the superexchange processes in the strong-coupling limit, the ones connecting the $t_{2g}$ states are expected to be antiferromagnetic, while the ones connecting the occupied $t_{2g}$ and unoccupied $e_g$ states will be ferromagnetic. Furthermore, there will be other contributions to $J_{1}$ beyond the strong-coupling limit~\cite{PRB2015}. Apparently, such FM and AFM contributions emerge in different ways in different computational schemes, which explains such a large spread in the values of $J_{1}$.

\par Next, let us consider the effect of the longer-range interactions $J_{2}$-$J_{6}$. Here, we will discuss only the $3d$$+$${\rm L}$ results, which take into account the effect of the ligand states. Generally, one can see that the longer-range interactions are ``more ferromagnetic'' in the case of CrI$_3$, while in CrCr$_3$ at least two interactions, $J_{2}$ and $J_{4}$, specifying the interlayer coupling, are always antiferromagnetic. The $\hat{m}$-scheme is an exception where all the interactions $J_{2}$-$J_{5}$ are antiferromagnetic, in both CrCl$_3$ and CrI$_3$. Then, for CrCl$_3$, the exchange parameters obtained within MFT correspond to the spin-spiral ground state with the propagation vector ${\bf k} = (\,0,\,0,\,0.69)$, in the units of reciprocal translations for the \emph{hexagonal} frame. In this spin texture, the spins in adjacent layers rotate relative to each other by nearly $90^{\circ}$. Very similar spin-spiral ground state with the propagation vectors ${\bf k} = (\,0,\,0,\,0.76)$ and $(\,0,\,0,\,0.72)$ is obtained in the $\hat{m}$- and $M$-based methods, respectively. For CrI$_3$, both MFT and $M$-based methods yield the FM ground state, in agreement with the experiment. On the contrary, the rotations of the spin magnetization matrix in the framework of the $\hat{m}$-scheme lead to the spin-spiral ground state with ${\bf k} = (\,0,\,0,\,0.84)$. The magnetic transition temperature for CrCl$_3$ and CrI$_3$, evaluated in RPA using the exchange parameters obtained in the framework of MFT and $M$-methods, are consistent with the experimental data ($T_{\rm X} =$ $17$ and $68$ K for CrCl$_3$ CrI$_3$, respectively~\cite{Bene1969}), while it is strongly overestimated in the $\hat{m}$-scheme.

\section{\label{sec:summary} Summary}
\par We have critically reexamined the problem of interatomic exchange interactions in SDFT or its refinements, where the ground-state magnetization is described by means of one-electron Kohn-Sham equations with some local (site-diagonal) xc potential. In this case, the interatomic exchange interactions can be associated with parameters of the Heisenberg model aiming to reproduce the total energy change of the real system caused by infinitesimal rotations of the magnetization near the ground state~\cite{LKG1984,LKG1985,LKAG1987}. Due to the perturbative character of the problem, such energy change can be always expressed in terms of the response function, which relates the change of the magnetization with the magnetic field inducing this change.

\par In the theory of exchange interactions, the input parameter is the magnetization change, which specifies the type of the perturbation near the ground state. Nevertheless, the magnetic field, which is required in order to produce this magnetization change can be formally obtained from the letter by means of the inverse response function. This constitutes the basis of the exact theory, where the exchange interactions are given by the inverse response function. Such theory should provide an exact estimate for the total energy change (at least within those approximations, which are typically additionally employed in SDFT).

\par In the context of the exchange interactions, the MFT relies on the additional assumption and, instead of using the response theory in order to find the required magnetic field, replaces it by the xc field corresponding to the input magnetization change. Although such identity holds for isolated atoms and follows from the general property of xc energy, it breaks down in solids, where the magnetization tends to additionally rotate towards the initial equilibrium state, being driven by the kinetic energy change. Then, although the exchange interactions in the framework of MFT can be still associated with the response function, this functional dependence appears to be linear. This is certainly an approximation, which affects the behavior of interatomic exchange interactions. Nevertheless, we would like to emphasize that, since MFT is based on the properties of xc field and energy, which become exact for isolated atoms, such theory is expected to work well in the strong-coupling limit~\cite{SClimit}.

\par We have studied these differences between MFT and the exact theory for the wide class of magnetic materials, including FM fcc Ni, AFM NiO, half-metallic FM CrO$_2$, multiferroic HoMnO$_3$, and layered van der Waals magnets CrCl$_3$ and CrI$_3$. We have argued that, although in a number of cases the MFT based approach provides quite a reasonable description on a semi-quantitative level, the exact theory is more consistent in several respects. Particularly, two important issues to be considered are: (i) the contributions of the ligand states, which under certain conditions can be eliminated by transferring their effects to the interaction parameters between the magnetic $3d$ states; and (ii) proper definition of the variable, which would describe the rotations of spins in SDFT. The first goal can be achieved by minimizing the magnetic energy change with respect to the ligand states for a given configuration of the $3d$ spins, as suggested by the adiabaticity concept, where the ``fast'' ligand degrees of freedom always follow the ``slow'' $3d$ spins. The second goal can be achieved also by minimizing the magnetic energy change, but with respect to the internal degrees of freedom, which describe the spin magnetization. By using this strategy, we have argued that the rotations of local spin moments are less energy costly, and therefore more suitable for the description of low-energy excitations, than the rotations of the full magnetization matrix. In order to describe properly all these effects, it is important do deal with the exact energy change, which is provided by the exact theory of exchange interactions.

\section*{Acknowledgement}
\par The work was supported by program AAAA-A18-118020190095-4 (Quantum).


\begin{thebibliography}{99}

\bibitem{PWA}
P. W. Anderson, Phys. Rev. \textbf{115}, 2 (1959).

\bibitem{Heisenberg}
W. Heisenberg,
Zeits. f. Physik \textbf{49}, 619 (1928).

\bibitem{RKKY}
M.~A. Ruderman and C. Kittel, Phys. Rev. \textbf{96}, 99 (1954);
T. Kasuya, Prog. Theor. Phys. \textbf{16}, 45 (1956);
K. Yosida, Phys. Rev. \textbf{106}, 893 (1957).

\bibitem{LKG1984}
A.~I. Liechtenstein, M.~I. Katsnelson, and V.~A. Gubanov, J. Phys. F: Met. Phys. \textbf{14}, L125 (1984).

\bibitem{LKG1985}
A.~I. Liechtenstein, M.~I. Katsnelson, and V.~A. Gubanov, Solid State Commun. \textbf{54}, 327 (1985).

\bibitem{LKAG1987}
A.~I. Liechtenstein, M.~I. Katsnelson, V.~P. Antropov, and V.~A. Gubanov, J. Magn. Magn. Mater. \textbf{67}, 65 (1987).

\bibitem{KL2004}
M.~I. Katsnelson and A.~I. Lichtenstein, J. Phys.: Condens. Matter \textbf{16}, 7439 (2004).

\bibitem{IS2003}
I.~V. Solovyev,
in \emph{Recent Res. Devel. Magnetism \& Magnetic Mat.}, Vol. 1 (Transworld Research Network, Kerala, India, 2003), pp 253-294, ISBN: 81-7895-100-2;
arXiv:cond-mat/0305668 (2003).

\bibitem{Kvashnin}
Y. O. Kvashnin, O. Gr{\aa}n{\"a}s, I. Di Marco, M. I. Katsnelson, A. I. Lichtenstein, and O. Eriksson,
Phys. Rev. B \textbf{91}, 125133 (2015).

\bibitem{Korotin}
Dm. M. Korotin, V. V. Mazurenko, V. I. Anisimov, and S. V. Streltsov,
Phys. Rev. B \textbf{91}, 224405 (2015).

\bibitem{Yoon}
H. Yoon, T.~J. Kim, J.-H. Sim, S.~W. Jang, T. Ozaki, and M.~J. Han,
Phys. Rev. B \textbf{97}, 125132 (2018).

\bibitem{Nomoto}
T. Nomoto, T. Koretsune, and R. Arita,
Phys. Rev. B \textbf{102}, 014444 (2020).

\bibitem{Grytsiuk}
S. Grytsiuk, J.-P. Hanke, M. Hoffmann, J. Bouaziz, O. Gomonay, G. Bihlmayer, S. Lounis, Y. Mokrousov, and S. Bl{\"u}gel,
Nature Communications \textbf{11}, 511 (2020).

\bibitem{AFT}
A.~K. Mackintosh and O.~K. Andersen, in \textit{Electrons at the Fermi
Surface}, edited by M. Springford (Cambridge University Press,
Cambridge, 1975); V. Heine, in \textit{Solid State Physics}, edited by H.
Ehrenreich, F. Seitz and D. Turnbull (Academic, New York,
1980), Vol. 35.

\bibitem{SClimit}
I.~V. Solovyev and K. Terakura,
Phys. Rev. Lett. \textbf{82}, 2959 (1999);
I.~V. Solovyev, Phys. Rev. B \textbf{60}, 8550 (1999).

\bibitem{Stocks}
G.~M. Stocks, B. Ujfalussy, X. Wang, D.~M.~C. Nicholson, W.~A. Shelton, Y. Wang, A. Canning, and B.~L. Gyorffy, Phil. Mag. \textbf{78}, 665 (1998).

\bibitem{BrunoPRL2003}
P. Bruno, Phys. Rev. Lett. \textbf{90}, 087205 (2003).

\bibitem{Antropov}
V.~P. Antropov, J. Magn. Magn. Mater. \textbf{262}, L192 (2003).

\bibitem{Savrasov}
S.~Y. Savrasov, Phys. Rev. Lett. \textbf{81}, 2570 (1998).

\bibitem{Grotheer}
O. Grotheer, C. Ederer, and M. F\"ahnle,
Phys. Rev. B \textbf{63}, 100401(R) (2001).

\bibitem{KeKatsnelson}
L. Ke and M.~I. Katsnelson, arXiv:2007.14518 [cond-mat.str-el].

\bibitem{HK}
P. Hohenberg and W. Kohn,
Phys. Rev. B \textbf{136}, B864 (1964).

\bibitem{KS}
W. Kohn and L.~J. Sham, Phys. Rev. A \textbf{140}, 1133 (1965).

\bibitem{WannierRevModPhys}
N. Marzari, A.~A. Mostofi, J.~R. Yates, I. Souza, and D.~Vanderbilt, Rev. Mod. Phys. \textbf{84}, 1419 (2012).

\bibitem{JPCMreview}
I.~V. Solovyev, J. Phys.: Condens. Matter \textbf{20}, 293201 (2008).

\bibitem{Vignale1987}
G. Vignale and M. Rasolt, Phys. Rev. Lett. \textbf{59}, 2360 (1987).

\bibitem{Vignale1988}
G. Vignale and M. Rasolt, Phys. Rev. B \textbf{37}, 10685 (1988).

\bibitem{PRB1998}
I.~V. Solovyev and K. Terakura, Phys. Rev. B \textbf{58}, 15496 (1998).

\bibitem{PRB2014}
I.~V. Solovyev, Phys. Rev. B \textbf{90}, 024417 (2014).

\bibitem{PRB2019}
O. Besbes, S. Nikolaev, N. Meskini, and I. Solovyev,
Phys. Rev. B \textbf{99}, 104432 (2019).

\bibitem{PCCP2019}
K. Wang, S. Nikolaev, W. Ren, and I. Solovyev,
Phys. Chem. Chem. Phys. \textbf{21}, 9597 (2019).


\bibitem{footnote1}
$\boldsymbol{\mathcal{R}}_{\boldsymbol{q}} = \frac{1}{2} \left(  \boldsymbol{\mathcal{R}}_{\boldsymbol{q}}^{\uparrow \downarrow} + \boldsymbol{\mathcal{R}}_{\boldsymbol{q}}^{\downarrow \uparrow} \right)$ in the notations of Ref.~\cite{PRB2014}.

\bibitem{Kanamori_GKA}
J. Kanamori, J. Phys. Chem. Solids \textbf{10}, 97 (1959).

\bibitem{spindynamics1}
V.~P. Antropov, M.~I. Katsnelson, B.~N. Harmon, M. van Schilfgaarde, and D. Kusnezov,
Phys. Rev. B \textbf{54}, 1019 (1996).

\bibitem{spindynamics2}
S.~V. Halilov, H. Eschrig, A. Y. Perlov, and P.~M. Oppeneer,
Phys. Rev. B \textbf{58}, 293 (1998).

\bibitem{Logemann}
R. Logemann, A.~N. Rudenko, M.~I. Katsnelson, and A. Kirilyuk,
J. Phys.: Condens. Matter \textbf{29}, 335801 (2017).

\bibitem{LMTO1}
O.~K. Andersen,
Phys. Rev. B \textbf{12}, 3060 (1975).

\bibitem{LMTO2}
O. Gunnarsson, O. Jepsen, and O.~K. Andersen,
Phys. Rev. B \textbf{27}, 7144 (1983).

\bibitem{AZA}
V.~I. Anisimov, J. Zaanen, and O.~K. Andersen,
Phys. Rev. B \textbf{44}, 943 (1991).

\bibitem{VWN}
S.~H. Vosko, L. Wilk, and M. Nusair,
Canadian Journal of Physics \textbf{58}, 1200 (1980).

\bibitem{tyab}
S.~V. Tyablikov, \emph{Methods of Quantum Theory of Magnetism}, Nauka, Moscow (1975).

\bibitem{fccNiexp}
H.~A. Mook and D.~McK. Paul,
Phys. Rev. Lett. \textbf{54}, 227 (1985).

\bibitem{Skomski}
R. Skomski, {\em Simple Models of Magnetism},
Oxford University Press, Oxford (2008).

\bibitem{Oguchi}
T. Oguchi, K. Terakura, and A.~R. Williams,
Phys. Rev. B \textbf{28}, 6443 (1983).

\bibitem{ZaanenSawatzky}
J. Zaanen and G.~A. Sawatzky,
Can. J. Phys. \textbf{65}, 1262 (1987).

\bibitem{NiOexp}
M.~T. Hutchings and E.~J. Samuelsen, Phys. Rev. B \textbf{6}, 3447 (1972).

\bibitem{PRM2019}
I.~V. Solovyev and S.~V. Streltsov,
Phys. Rev. Materials \textbf{3}, 114402 (2019).

\bibitem{Mazin}
I.~I. Mazin, D.~J. Singh, and C. Ambrosch-Draxl,
Phys. Rev. B \textbf{59}, 411 (1999).

\bibitem{HMRevModPhys}
M.~I. Katsnelson, V.~Yu. Irkhin, L. Chioncel, A.~I. Lichtenstein, and R.~A. de Groot,
Rev. Mod. Phys. \textbf{80}, 315 (2008).

\bibitem{CrO2ARPES}
F. Bisti, V.~A. Rogalev, M. Karolak, S. Paul, A. Gupta, T. Schmitt, G. G\"untherodt, V. Eyert, G. Sangiovanni, G. Profeta, and V.~N. Strocov,
Phys. Rev. X \textbf{7}, 041067 (2017).

\bibitem{Porta}
P. Porta, M. Marezio, J.~P. Remeika, and P.~D. Dernier,
Mater. Res. Bull. \textbf{7}, 157 (1972).

\bibitem{PRB2015}
I.~V. Solovyev, I.~V. Kashin, and V.~V. Mazurenko,
Phys. Rev. B \textbf{92}, 144407 (2015).

\bibitem{JPCM2016}
I.~V. Solovyev, I.~V. Kashin, and V.~V. Mazurenko,
J. Phys.: Condens. Matter \textbf{28}, 216001 (2016).

\bibitem{ExpStructureHoMnO3}
A. Mu\~noz, M.~T. Cas\'ais, J.~A. Alonso, M.~J. Mart\'inez-Lope,
J.~L. Mart\'inez, and M.~T. Fern\'andez-D\'iaz,
Inorg. Chem. \textbf{40}, 1020 (2001).

\bibitem{Ishiwata}
S. Ishiwata, Y. Kaneko, Y. Tokunaga, Y. Taguchi, T.~H. Arima, and Y. Tokura,
Phys. Rev. B \textbf{81}, 100411(R) (2010).

\bibitem{Picozzi}
S. Picozzi, K. Yamauchi, B. Sanyal, I.~A. Sergienko,
and E. Dagotto, Phys. Rev. Lett. \textbf{99}, 227201 (2007).

\bibitem{Okuyama}
D. Okuyama, S. Ishiwata, Y. Takahashi, K. Yamauchi, S. Picozzi, K. Sugimoto, H. Sakai,
M. Takata, R. Shimano, Y. Taguchi, T. Arima, and Y. Tokura,
Phys. Rev. B \textbf{84}, 054440 (2011).

\bibitem{PRB11}
I.~V. Solovyev,
Phys. Rev. B \textbf{83}, 054404 (2011); \textit{ibid.} \textbf{90}, 179910 (2014).

\bibitem{PRB12}
I.~V. Solovyev, M.~V. Valentyuk, and V.~V. Mazurenko,
Phys. Rev. B \textbf{86}, 144406 (2012).

\bibitem{JPSJ}
I. Solovyev, J. Phys. Soc. Jpn. \textbf{78}, 054710 (2009).

\bibitem{Heine}
V. Heine and J.~H. Samson, J. Phys. F: Metal Phys. \textbf{10}, 2609 (1980); \textit{ibid.} \textbf{13}, 2155 (1983).

\bibitem{CrI3_Nature}
B. Huang, G. Clark, E. Navarro-Moratalla, D.~R. Klein, R. Cheng, K.~L. Seyler, D. Zhong, E. Schmidgall, M.~A. McGuire, D.~H. Cobden, W. Yao, D. Xiao, P. Jarillo-Herrero, and X. Xu,
Nature (London) \textbf{546}, 270 (2017).

\bibitem{CrCl3str}
B. Morosin and A. Narath,
J. Chem. Phys. \textbf{40}, 1958 (1964).

\bibitem{CrI3str}
M.~A. McGuire, H. Dixit, V.~R. Cooper, and B.~C. Sales,
Chemistry of Materials \textbf{27}, 612 (2015).

\bibitem{Bene1969}
R.~W. Ben\'e, Phys. Rev. \textbf{178}, 497 (1969).

\end{thebibliography}
\end{document}